\documentclass[12pt]{article}
\usepackage{amsmath}
\usepackage{graphicx,psfrag,epsf}
\usepackage{enumerate}
\usepackage{natbib}
\usepackage{url} 
\usepackage{chngcntr}
\usepackage{amssymb}
\usepackage{float}
\usepackage[caption = false]{subfig}

\newtheorem{Def}{Definition}

\newtheorem{Prop}{Proposition}
\DeclareMathOperator*{\arginf}{arg\,inf}

\newcommand{\blind}{0}

\usepackage{setspace}

\begin{document}

\def\spacingset#1{\renewcommand{\baselinestretch}%
{#1}\small\normalsize} \spacingset{1}


\if0\blind
{
  \title{\bf On perfect sampling: ROCFTP with Metropolis-multishift coupler}
  \author{Nabipoor, Majid \\ 
   \texttt{nabipoor@ualberta.ca}\\   
    University Health Network,\\
    Toronto, Canada}
  \maketitle
} \fi

\if1\blind
{
  \bigskip
  \bigskip
  \bigskip
  \begin{center}
    {\LARGE\bf On perfect sampling: ROCFTP with Metropolis-multishift coupler}
\end{center}
  \medskip
} \fi

\bigskip
\begin{abstract}
ROCFTP is a perfect sampling algorithm that employs various random operations, and requiring a specific Markov chain construction for each target. To overcome this requirement, the Metropolis algorithm is incorporated as a random operation within ROCFTP. While the Metropolis sampler functions as a random operation, it isn't a coupler. However, by employing normal multishift coupler as a symmetric proposal for Metropolis, we obtain ROCFTP with Metropolis-multishift. Initially designed for bounded state spaces, ROCFTP's applicability to targets with unbounded state spaces is extended through the introduction of the Most Interest Range (MIR) for practical use. It was demonstrated that selecting MIR decreases the likelihood of ROCFTP hitting $MIR^C$ by a factor of $(1 - \varepsilon)$, which is beneficial for practical implementation. The algorithm exhibits a convergence rate characterized by exponential decay. Its performance is rigorously evaluated across various targets, and tests ensure its goodness of fit. Lastly, an R package is provided for generating exact samples using ROCFTP Metropolis-multishift.  
\end{abstract}

\noindent%
\vfill

\doublespacing

\section{Introduction}
\label{sec:intro}

Many applications of Markov Chain Monte Carlo (MCMC) are centered around Bayesian inference. Bayesian inference relies on posterior features such as moments, quantiles, or the highest posterior density region, which can be expressed in terms of posterior expectations of functions of $\theta$, denoted as $E[f(\theta) | Y]$, where $\theta$ is a model parameter, and $Y$ represents the observed data.

The integration in this expression has been a challenging aspect of Bayesian inference, and in most applications, the analytic evaluation of $E[f(\theta) | Y]$ is not possible. Alternative approaches include: (1) Numerical evaluation, which can be difficult and occasionally inaccurate (2) Analytic approximation methods like Laplace approximation, which are sometimes suitable (3) Monte Carlo integration, including MCMC. The classic approach of MCMC involves evaluating $E[f(X)]$ by drawing samples ${X_t: t=1, ..., n}$ from the distribution of $X$ and approximating it as:
$E[f(X)] \approx \frac{1}{n} \sum_{t=1}^n f(X_t).$ 
This equation represents an ergodic average, and convergence to the required expectation is guaranteed by the strong ergodic theorem. However, obtaining samples ${X_t}$ can be challenging since posteriors are often non-standard. One approach to address this is through a Markov chain with $\pi(\cdot)$ as its stationary distribution. 

The ROCFTP algorithm, proposed by \protect{\cite{wilson2000couple}}, offers a coupling diagnostic approach for drawing independent and identically distributed (i.i.d.) samples from a Markov Chain. However, it requires the construction of a Markov chain for each target and can be executed using various random operation couplers. The Metropolis sampler functions as a random operation but not as a coupler, whereas the normal multishift serves as a symmetric coupler that can be employed as a proposal in the Metropolis algorithm. ROCFTP with Metropolis-multishift coupler can generate i.i.d. samples without the necessity of constructing a Markov chain for each target.

This manuscript introduces the perfect sampling algorithm of ROCFTP with Metropolis-multishift coupler and investigates its properties thoroughly. The associated R package $ROCFTP.MMS$ is available on the Comprehensive R Archive Network (CRAN) at \emph{https://cran.r-project.org/web/packages/ROCFTP.MMS/index.html}. Section 2 delves into the fundamentals and definitions of random operations, the normal multishift coupler, CFTP, and ROCFTP, while also addressing the determination of the block length $T$ in ROCFTP. Section 3 further explores the Metropolis-multishift coupler, including its functionality, convergence properties, the impact of starting range selection on sampling, and considerations regarding the number of starting paths. In Section 4, the performance of ROCFTP with the Metropolis-multishift coupler is assessed on both unimodal and multimodal targets. Lastly, Section 5 provides a detailed discussion of the results, along with a  comparison with other algorithms.

\subsection{Literature review}
The convergence of an MCMC algorithm to its stationary distribution is a necessary condition for drawing inferential conclusion of MCMC runs, \protect{\cite{frigessi1995bayesian}}. Convergence properties of MCMC algorithms have been widely studied for decades; \protect{\cite{peskun1973optimum, sokal1988absence, besag1993spatial, frigessi1995bayesian, mengersen1996rates, johnson1998coupling}}. One way of drawing samples in MCMC algorithms is to discard the first m iterations, or draw samples after a sufficiently long burn-in of say m iterations to allow Markov chain converges to its stationary distribution \protect{\cite{besag1995bayesian}}. \protect{\cite{frigessi1995bayesian}} defines the burn-in $t^*$ such that for all $t> t^*, \; \; \|P(X^{(t)}= \cdot \; | \; x^{(1)})-\pi(\cdot)\| \; \leq \; \varepsilon$, and investigates different methods to determine upper and lower bounds for $t^*$. In this approach, sample points are dependent, \protect{\cite{mykland1995regeneration}}. As a potential remedy, it is suggested that many independent runs of MCMC be used; however, it is expensive, \protect{\cite{besag1993spatial}}. The other approach is sub-sampling the chain at equally spaced times after a sufficiently long burn-in, \protect{\cite{besag1993spatial}}. Regenerative simulation and splitting Markov chain are two different proposed techniques for sub-sampling from a chain. \protect{\cite{mykland1995regeneration}} assumed $T_0 \leq T_1 \leq \dots$ such that at each $T_i$, the future of the process is independent of the past and identically distributed; and proposed a \emph{scaled regeneration quantile} plot to determine such $T_i$s, which are known as regeneration times. Consider a discrete Markov chain, and fix a particular state, the times that chain returns to that state are regeneration times. \protect{\cite{nummelin1978splitting}} generalized this idea to Markov chains with continuous state spaces by introducing the concept of an \emph{atom}. \protect{\cite{nummelin1978splitting}} defines an \emph{atom} as a set of points of which all transitions are identical, and the chain is split by entering to the \emph{atom}.

Coupling is the joint construction of two or more random variables or Markov chains, and as a probabilistic technique dates back to \protect{\cite{doeblin1938expose}}. Later, this technique was used to prove ergodic theorems for Markov lattice interactions, \protect{\cite{wasserstein1969markov}}, \protect{\cite{dobrushin1971markov}}, \protect{\cite{holley1975ergodic}}. Advancements in coupling theory led to the development of various coupling schemes, including quantile coupling, maximal coupling, Ornstein coupling, epsilon-coupling, and shift coupling. Maximal coupling which attains the upper bound of coupling event inequality proposed by \protect{\cite{goldstein1979maximal}}, was later utilized for poison approximation by \protect{\cite{thorisson1995coupling}}. \protect{\cite{lindvall1992lectures}} compiled the scattered literature on coupling schemes, laying the groundwork for \protect{\cite{thorisson2000appl}}'s book. \protect{\cite{letac1986contraction}} observed that if a Markov chain moves backward in time, it will pointwise converge to the stationary distribution. \protect{\cite{aldous1990random}} investigated a Markov chain with random walk stationary, and through a reversed time transition matrix, proved that the chain converges to uniform distribution. \protect{\cite{johnson1996studying}} proposed monotone coupling for a $p$-dimensional Markov chain where all conditional distributions are monotonically decreasing outside a $p$-dimensional hypercube. A novel approach, proposed by \protect{\cite{propp1996exact}}, Coupling From The Past (CFTP), involves running two or more chains in parallel until all chains coalesce. This algorithm outputs an exact draw from the stationary distribution.  Monotone CFTP looks backward to find the two chains that coalesce in a block and the state at time 0 is according with the stationary distribution $\pi$. The selection of random operations in CFTP is as crucial as the choice of the Markov chain itself. Monotone random operations are particularly important, as they enable the algorithm to confine the entire state space to a single point. The CFTP algorithm necessitates the storage of randomness sources, rendering it costly. \protect{\cite{wilson2000couple}} proposed Read-Once CFTP (ROCFTP), which operates Markov chains in a forward manner. The algorithm reads a stream of randomness only once, leading to significant reductions in memory and time requirements. Later, \protect{\cite{wilson2000layered}} explored various monotone random operations, with a focus on uniform multishift and normal multishift. CFTP is applicable on bounded state spaces, with $\hat{0}, \hat{1}$ representing the unique minimum and maximum of the state space, respectively. The algorithm initiates at time $-t$ in the past with initial values of $\hat{0}, \hat{1}$. If these two chains coalesce in a block, the observation at time zero serves as the output. Fill's algorithm, introduced by \protect{\cite{fill1997interruptible}}, is structured in a manner similar to CFTP. Fill's algorithm initiates a chain $X$ at time zero from $X_0=\hat{0}$ and advances the chain forward for $t$ steps by storing the randomness source. Then, at time $t$, the second chain $Y$ starts from $Y_t=\hat{1}$ and runs backward based on the stored randomness. If $Y_0=\hat{0}$, then the algorithm outputs $X_t$, if not, it doubles $t$, and repeats the process. Fill explained that every iteration of doubling $t$ implies rejection sampling and investigated the reasons behind the effectiveness of this algorithm. However, this algorithm can be interrupted with respect to a deadline specified in terms of Markov chain steps. Fill's algorithm was extended by \protect{\cite{fill2000extension}}, making it applicable to general chains, and capable of alleviating computational burdens. The algorithm initiates a chain at time $t$ with initial value of $X_t$ and runs the time-reversed chain for $t$ steps to obtain $X_0$, while stores the randomness. Then, at time zero, it starts a second chain $Y$ with an arbitrary initial value from the state space and runs forward according to the stored randomness. If both chains coalesce at time $t$, it outputs $X_0$, otherwise, it repeats the algorithm with a new $t$ and $X_t$. Fill's algorithm outputs $X_t$, while the extended version outputs $X_0$. \protect{\cite{berthelsen2010perfect}} used ROCFTP with Gibbs sampler to generate samples for mixture models with unknown mixture weights.

\protect{\cite{glynn2014exact}} proposed a new approach for unbiased estimation of Markov chain stationary expectations. They defined $Z:=\sum_{K=0}^{N}\frac{\Delta_k}{P(N \ge k)}$ as an estimator of $E\left[f(X)\right]$ where $\Delta_k=f(X_k)-f(X_{k-1})$, and $N$ is a $\mathbb{Z}_+$-valued random variable independent of $\Delta_k$. They established a coupling between $X_k$ and $X_{k-1}$ to ensure $\Delta_k \longrightarrow 0$ as $k \longrightarrow \infty$. They demonstrated that under three conditions of (1) Markov chain $X$ is counteractive on average, (2) $f$ is Lipschitz, (3) the jump in each transition of the Markov chain $X$ is finite; then $Z$ is an unbiased estimator for $E\left[f(X)\right]$.  \protect{\cite{jacob2020unbiased}} employed the telescopic sum argument of Glynn-Rhee combined with the coupling of Markov chains to derive an unbiased estimate for $E\left[f(X)\right]$. Additionally, \protect{\cite{leigh2023perfect}} demonstrated that unbiased simulation, coupled with the coupling of two Markov chains, achieves perfect simulation.

\section{Preliminaries}
This section provides preliminaries, including random operations, normal multishift coupler, CFTP and ROCFTP for constructing ROCFTP with Metropolis-multishift. Advanced reader may skim through this section; however, subsection 2.4 is recommended, as it provides a simulation for ROCFTP to investigate the length of the block $T$. The scheme of storing randomness in CFTP algorithm, along with related R-code, is provided in the Appendix,  laying the foundation for further materials.   

\subsection{Random operations}
Random operation $\phi()$ is a deterministic function that takes the input state $X_t$ at time $t$ with some intrinsic randomness $U_{t+1}$; updates the state $X_t$, and outputs $X_{t+1}=\phi(X_t,U_{t+1})$, \protect{\cite{wilson2000layered}}. The intrinsic randomness $U_t$s are mutually independent. The random operation $\phi()$ preserves the stationary distribution $\pi$; if $X_t$ is distributed according to $\pi$ and $U_t$ is random, then $X_{t+1}$ is distributed according to $\pi$. 
\begin{Def}
A random operation $\phi()$ is monotone if:
$$X_t \leq Y_t \Longrightarrow \phi(X_t,U_{t+1}) \leq  \phi(Y_t,U_{t+1}).$$
\end{Def}
Monotonicity is a useful property for sandwiching the entire state space at time $t$ in a state $X_t$, \protect{\cite{wilson2000layered, murdoch2000exact}}. Consider a finitely bounded continuous state space and denote $\hat{0}$ as minimum and $\hat{1}$ as maximum of the state space. In order to test if the random operation $\phi()$ maps simultaneously the entire state space in $X_t$; by monotonicity, it is necessary only to apply it on the two starting values $\hat{0}$ and $\hat{1}$. A multishift coupler is a random operation $\phi()$, such that for each real value $x$, the random value $\phi(x,u)-x$ has the same target distribution \protect{\citep{wilson2000layered}}. In other words, the update distribution  belongs to a location family. There are different multishift couplers, such as uniform multishift coupler, normal multishift coupler, and unimodal multishift coupler. We will discuss the normal multishift coupler.

\subsection{Normal multishift coupler}
\begin{figure}[ht]
\begin{center}
\includegraphics[width=4.5in]{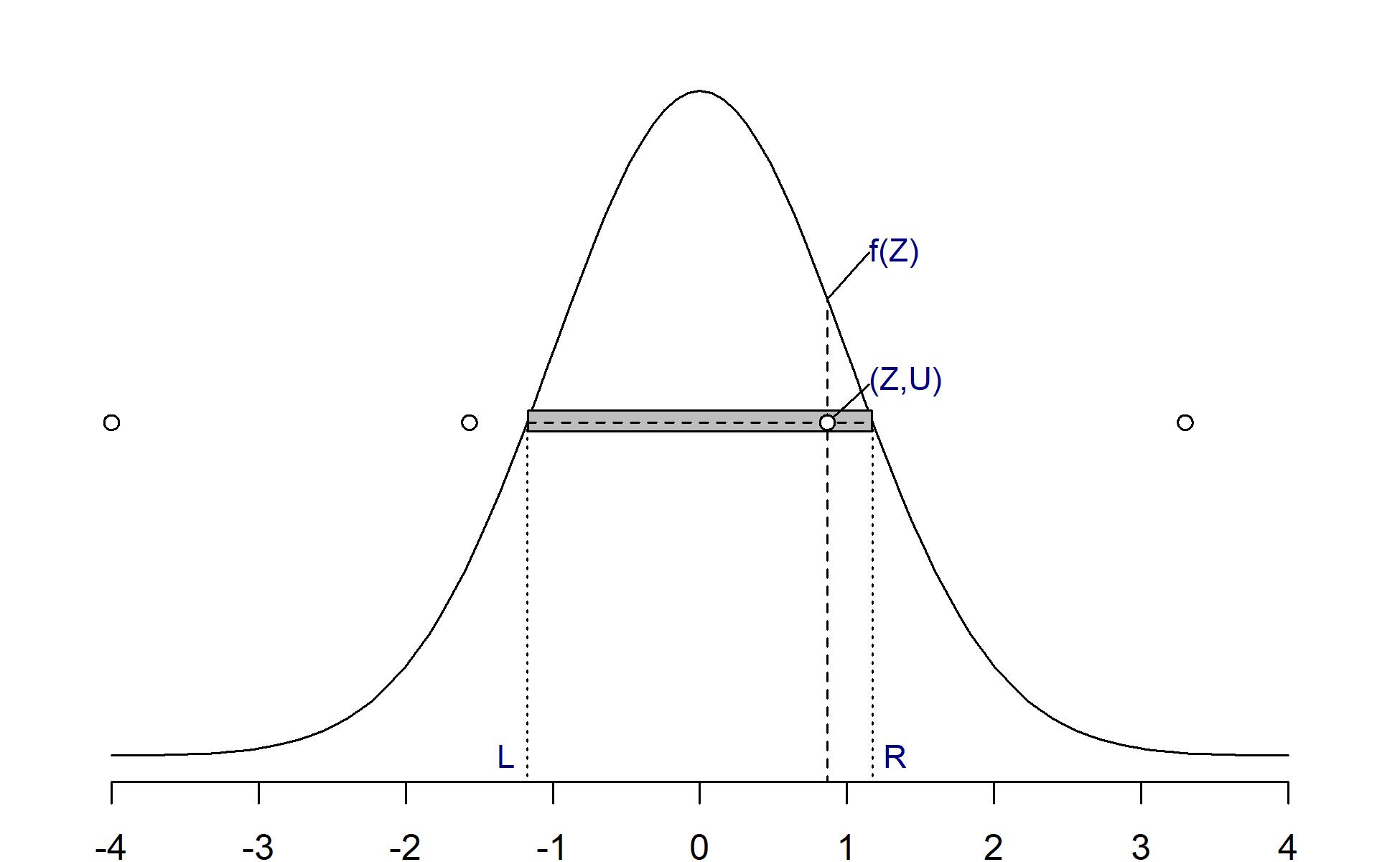}
\end{center}
\caption{Illustration of normal multishift coupler. \label{fig:2}}
\end{figure}

\protect{\cite{wilson2000layered}} proposed the normal multishift coupler, and explained it as a combination of layered uniform rectangles, as depicted in Figure \ref{fig:1}. He referred to it as the layered normal multishift coupler. In normal multishift coupler, we select a rectangle from the mixture of uniform distributions according to the mixing proportions, and then pick a random point within the rectangle. The result is normally distributed random variable. Let $L$ and $R$ denote the left and right end points of a uniform rectangle, and let $f$ be the density of normal distribution. Choose $Z \sim Normal(0,1)$ and $U \sim Uniform (0,f(Z))$. Therefore $L=-f^{-1}(U)$ and $R=f^{-1}(U)$, where $f^{-1}$ is the inverse function of $f$. Take a random point within the rectangle $X \sim Uniform(L,R)$, and apply mapping $ M_{L,R,X}(s)=\lfloor \frac{s+R-X}{R-L} \rfloor (R-L) + X$. This method is illustrated in Figure \ref{fig:2}.
In this algorithm, $\sigma =1$, and it can be changed for any $\sigma$. Note that $s$ is the initial value or the previous state. In fact, if we define a Markov chain, $s$ will change based on every update of the chain. For instance, let $X_{t+1}=\rho X_t + \varepsilon_{t+1}$, where $\rho$ is given. Take $s=\rho X_t$ in the mapping and repeat the procedure. 

\subsection{CFTP}
\begin{figure}[ht]
\begin{center}
\includegraphics[width=4in]{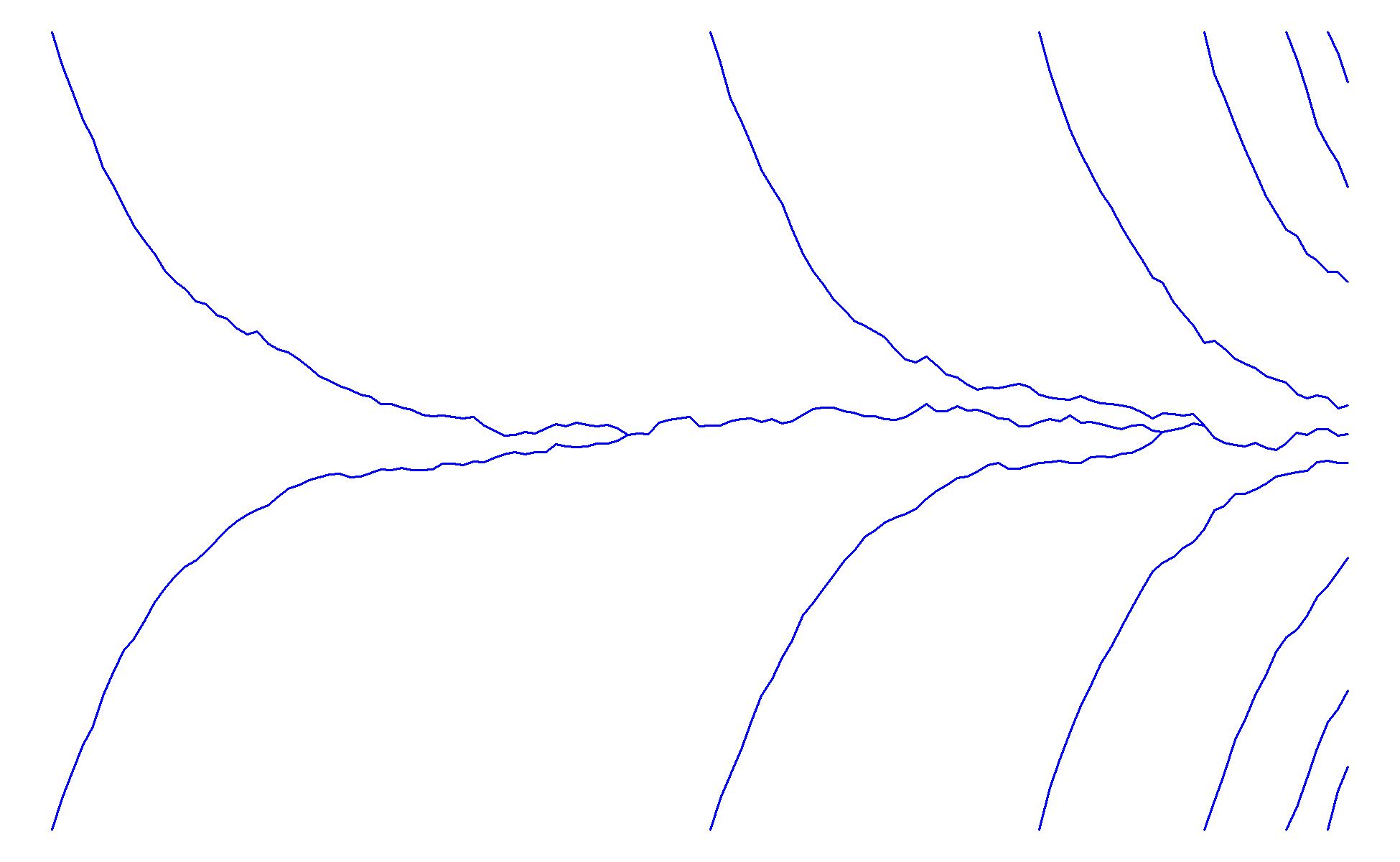}
\end{center}
\caption{Illustration of CFTP in monotone case with normal multishift coupler. \label{fig:3}}
\end{figure}

Consider an ergodic discrete Markov chain with transition probability $p_{ij}>0$ from state $i$ to state $j$; ergodicity implies that there exists a unique stationary probability distribution $\pi(\cdot)$. Let, the Markov chain start in some state and run for a long time; the probability that it ends up in state $i$ converges to $\pi(i)$. CFTP is the extreme implementation of this principle, considering all possible starting values simultaneously, and running the chains until dependence on the starting values has vanished. Consider a collection of coupled Markov chains started simultaneously at every possible state of the chain at time $t=-2^n$ and end at time $t=0$. If the paths occasionally coalesce into one path in the block of $(-2^n, -2^{n-1})$, then we are done; if not, double $t$ and start again, by using the stored sequence of randomness. CFTP is a simulation algorithm that can be executed using various random operations. If the random operation is  monotone, it allows the algorithm to sandwich all the state space in one state. Hence, it suffices to start only two chains from $\hat{0}$ and $\hat{1}$.  In fact, CFTP looks backward block by block (doubles each time) until the two chains of $\hat{0}$ and $\hat{1}$ started at $t=-2^n$ coalesce on $n$th block and outputs $X_0$. Suppose $\{X_t\}$ is a Markov chain with state space $\mathcal{X}$; $\phi()$ is a random operation $X_{t}=\phi (X_{t-1}, U_{t})$, and $U_t$ is the randomness at time $t$. Then CFTP algorithm is:

\begin{quote}
    $CFTP(2^n): $\\
    $t \longleftarrow  -2^n $\\
    $B_{t}\longleftarrow \mathcal{X} $\\
    While $\;  t<0$ \\
    $ \; t\longleftarrow t+1 $\\
    $ \; B_{t}\longleftarrow \phi (B_{t-1}, U_{t}) $\\
    If $\; | B_{-2^{n-1}}|=1 \ $ \\
    $ \;   $ Return$(B_{0}) $\\
    Else  \\
    $ \; CFTP(2^{n+1}) $
\end{quote}

where $|\cdot|$ denotes the cardinality of a set. CFTP is based on the observation that if the chain was run from all initial states in $\mathcal{X}$ at time $t=-\infty, \; (B_{-\infty}=\mathcal{X})$, all paths will coalesce to a singleton at time $t=0, \; (B_0=\{X_0\})$, and $X_0$ has the stationary distribution $\pi$. Figure \ref{fig:3} illustrates CFTP in monotone case with a normal multishift coupler. The trajectories have the same starting points but in different times. The interested reader is refereed to \protect{\cite{huber2016perfect}} for more detail.
It is useful to describe how to illustrate CFTP because it explains how and why this algorithm stores the randomness $U_t$. Figure \ref{fig:4} shows how to construct the matrix of CFTP to store randomness. Suppose we have a function, let's call it multishift(), which, given a random vector, generates a new random vector. In fact, it takes the initial values and produces the start block in Figure \ref{fig:4}, increasing it by $2^n$ where $n=2,3,\cdots$ in the next iterations. The second block contains $2^2$ new vectors, shown as a white block in the top right of the start block. It is necessary to continue this block to time $t=0$. Note that this sampling is simultaneous, so we have to store the randomness $U_t$ which is used in the start block. Then add the NA's block as shown in Figure \ref{fig:4}, and continue the process recursively until coalescence in a block. The R code for Figure \ref{fig:3} is provided in Appendix \ref{CFTP code}.

\subsection{ROCFTP}
\begin{figure}[ht]
\begin{center}
\includegraphics[width=5in]{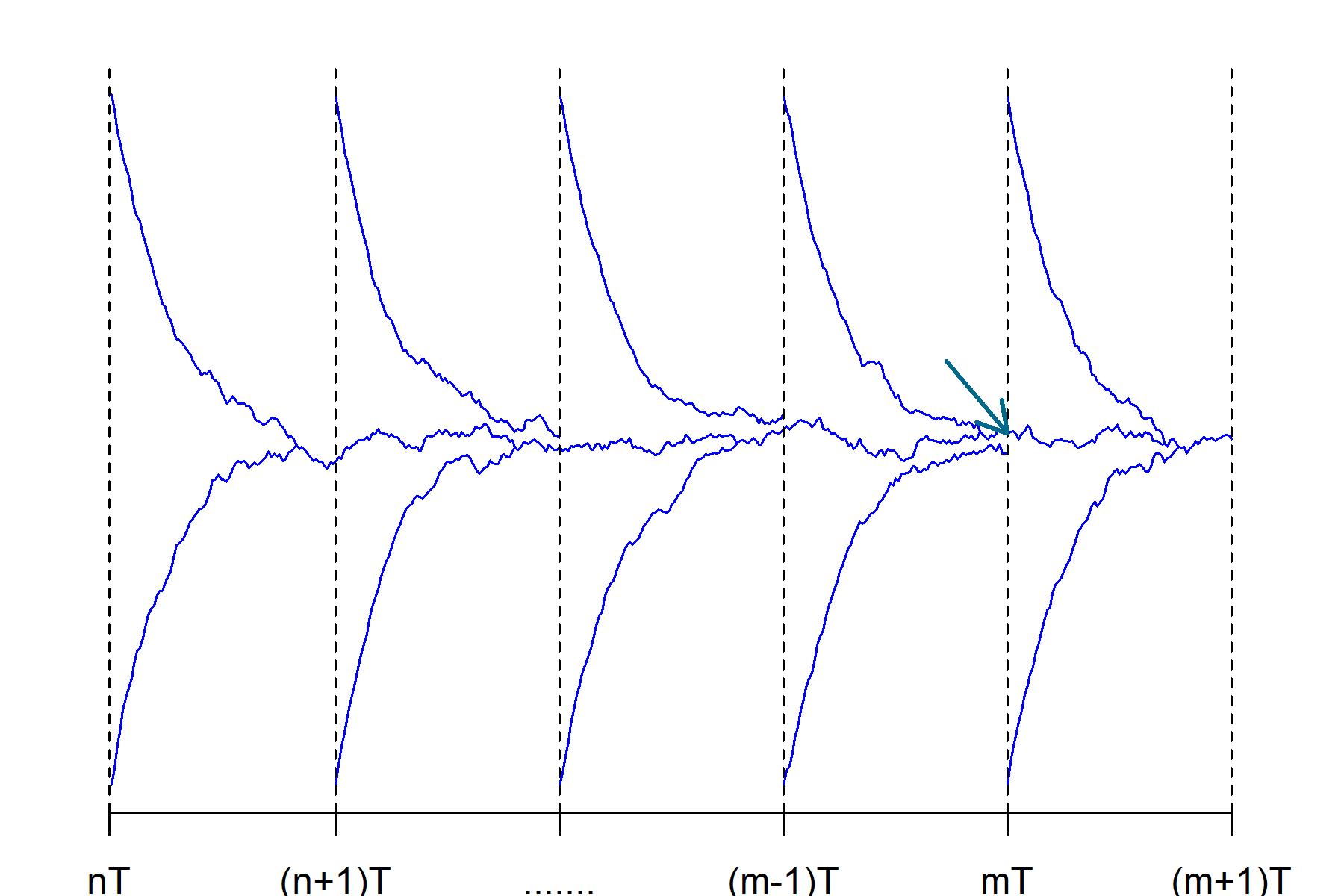}
\end{center}
\caption{Illustration of ROCFTP: output is $X_{mT}$. \label{fig:5}}
\end{figure}

CFTP by increasing the length of block from $M$ to $2M$ requires $U_{-2M}, \cdots, U_0$ in such a way that $U_{-M+1},\cdots,U_{0}$ from the previous blocks must not change. It is needed to store randomness from $U_{-M+1}$ to $U_{0}$, and not only this storing is difficult to implement, but also it requires lots of  storage space, specifically when the random operation is not monotone. These issues led to the development of ROCFTP, proposed by \protect{\cite{wilson2000couple}}, which employs a retroactive stopping rule. It involves executing random operations in a forward direction over time. Then, at a certain point, it decides to halt and return a previous state instead of continuing from the current state. The time notation in ROCFTP is an inner notation and it uses each $U_t$ once without any storage. Let $C_{-n}= \{\textnormal{paths coalesce in block } [-nT,-(n-1)T]\}$, therefore, the pairs $(C_{-n},X_{-(n-1)T})$ depend only on $\{U_t\}_{t=-nT+1}^{-(n-1)T}$, so such pairs are independent across $n$. In Figure \ref{fig:5}, suppose CFTP runs at $mT$ as the present time. It means that at least one $C_{-n}$ event, $n=1,\cdots,m$ has occurred. Then, we could search backward for the first previous $C_{-n}$ event. The output will be the update at the present time or $t=mT$. This means, we could relabel the time axis, and $X_{mT}$ could relabel as $X_0$, which is the output of CFTP. This is a simpler perfect sampler, and we don't need to reuse $U_t$ values, it only requires saving $X_{mT}$ values as output. 
The algorithm of ROCFTP is as follows: 
\begin{quote}
    1. Run from $t=0$ forward to the first $C_n$ event.\\
    2. Follow the paths from $X_{(n+1)T}$; and run from $t=(n+1)T$ until $C_m$, where $m\geq n$, occurs.\\
    3. Output $X_{mT}$.
\end{quote}
With this algorithm, one question arises. What is $T$, and how do we choose it? Suppose we can choose $T\leq t$ and a coupling such that
$$P(\textmd{coalescence in the first block }|\; T)=p.$$
Blocks run independently, therefore:
$$P(1^{st} \textmd{ coalescence in the block n } |\; T)=p(1-p)^{n-1}.$$
This is a geometric random variable, so $E[n\;|\;T]=1/p$. Suppose $\tau$ is the total time that ROCFTP needs to generate one sample. Therefore $\tau=nT$ and by considering $T\leq t$:
$$E[\tau]=E[nT]=E\left[E[nT\;|\;T]\right] \leq \frac{t}{p}.$$

A simulation conducted to explore the effects of the block length $T$ in the ROCFTP algorithm. Let $p$ denote the probability of coalescence in the first block, $n$ represent the number of blocks needed until the first block with coalescence, and $\tau$ indicate the total time to generate a sample. We implemented a ROCFTP with mutishift coupler and target distribution $N(0,1)$, and executed the algorithm $100K$ times for starting ranges of $(-100, 100)$, $(-50,50)$, $(-10, 10)$, with various selections for block time $T$. It is evident that as $T$ increases,  $p$ also increases, and the total time  $\tau$ to generate a sample decreases up to a certain range, beyond which $\tau$ starts to increase. For more details, refer to Table \ref{T1}.\\

\begin{table}[ht]
\begin{center}
\begin{tabular}{lccccccccc}
\hline \hline
$(\hat{0}, \hat{1})$  & $T$ & $p$   & $\Bar{n}$ & $\Bar{\tau}$ & $(\hat{0}, \hat{1})$ & $T$ & $p$  & $\Bar{n}$ & $\Bar{\tau}$\\ \hline
(-100,100) & 70  & 0.244 & 4.113 & 288 & (-50,50) & 50  & 0.102 & 9.710 & 485 \\
           & 80  & 0.481 & 2.062 & 165 &          & 60  & 0.327 & 3.051 & 183 \\
           & 90  & 0.674 & 1.481 & 133 &          & 70  & 0.556 & 1.794 & 126 \\
           & 100 & 0.800 & 1.249 & 125 &          & 80  & 0.725 & 1.380 & 110 \\
           & 110 & 0.880 & 1.137 & 125 &          & 90  & 0.833 & 1.200 & 108 \\
           & 120 & 0.930 & 1.075 & 129 &          & 100 & 0.901 & 1.109 & 111 \\ \hline
(-10,10)   & 20  & 0.107 & 9.395 & 188 &          &     &       &       &     \\
           & 30  & 0.356 & 2.821 & 85  &          &     &       &       &     \\
           & 40  & 0.585 & 1.712 & 68  &          &     &       &       &     \\
           & 50  & 0.745 & 1.343 & 67  &          &     &       &       &     \\
           & 60  & 0.844 & 1.185 & 71  &          &     &       &       &     \\ \hline \hline
\end{tabular}
\end{center}
\caption{Simulation for the time block $T$ by running 100k iteration of ROCFTP with a normal mutishift coupler and target distribution $N(0,1)$. The total time $\tau$ decreases as $T$ increases up to a certain range. \label{T1}}
\end{table}

In CFTP or ROCFTP, once coalescence occurs, the past is forgotten, and no initialization bias remains. However, removing the initialization bias alone may not suffice, as there could still be a coupling time bias. The running time $T$ of CFTP is an unbounded random variable, which is unknown a priori.
\protect{\cite{fill1997interruptible}} noted that if we terminate the process when $T>T^*$, we introduce the {\it impatience bias}. In such cases, we are more likely to observe outcomes resulting from fast coalescence. For instance, consider an asymmetric random walk on $\{1,\cdots,N\}$ that steps up 1, down 2. It can coalesce at 1 in $N/2$ steps, but requires at least $N-1$ steps to coalesce at $N$. To address this, he introduced a variation of ROCFTP that does not exhibit impatience bias, i.e., it is interruptible. Latter, \protect{\cite{fill2000extension}} extended Fill's algorithm to general chains on quite general state spaces. In fact, they showed that given coalescence, the value at the end of the block is independent of the value at the start.

\section{Metropolis-multishift coupler}
In the previous section, we covered CFTP and ROCFTP, both employing different random operations. While the multishift coupler offers monotonicity, ideal for encompassing the entire state space, it lacks a built-in Markov chain. Establishing a suitable Markov chain for the target distribution becomes essential. On the other hand, traditional methods like the Metropolis algorithm don't demand this construction but lack monotonicity. Consequently, they cannot encapsulate the entire state space in one point.

In the Metropolis sampler, we have the freedom to choose the proposal distribution. The algorithm is theoretically guaranteed to converge to its stationary distribution regardless of the proposal distribution selected. However, in practice, the choice of proposal distribution is critical as it significantly impacts the convergence time. While the Metropolis algorithm utilizing a normal multishift proposal tends to drive proposed values toward the target distribution in a monotonically convergent manner, the acceptance-rejection mechanism within the algorithm prevents strict monotonicity in the generated chain. Nevertheless, owing to the normal multishift proposal, the chain typically demonstrates behavior closely resembling monotonicity. We will begin by developing and applying the Metropolis-multishift algorithm to various target distributions before integrating it as a random operation in ROCFTP.

\subsection{Metropolis sampler}
\begin{figure}[ht]
\begin{center}
\includegraphics[width=4in]{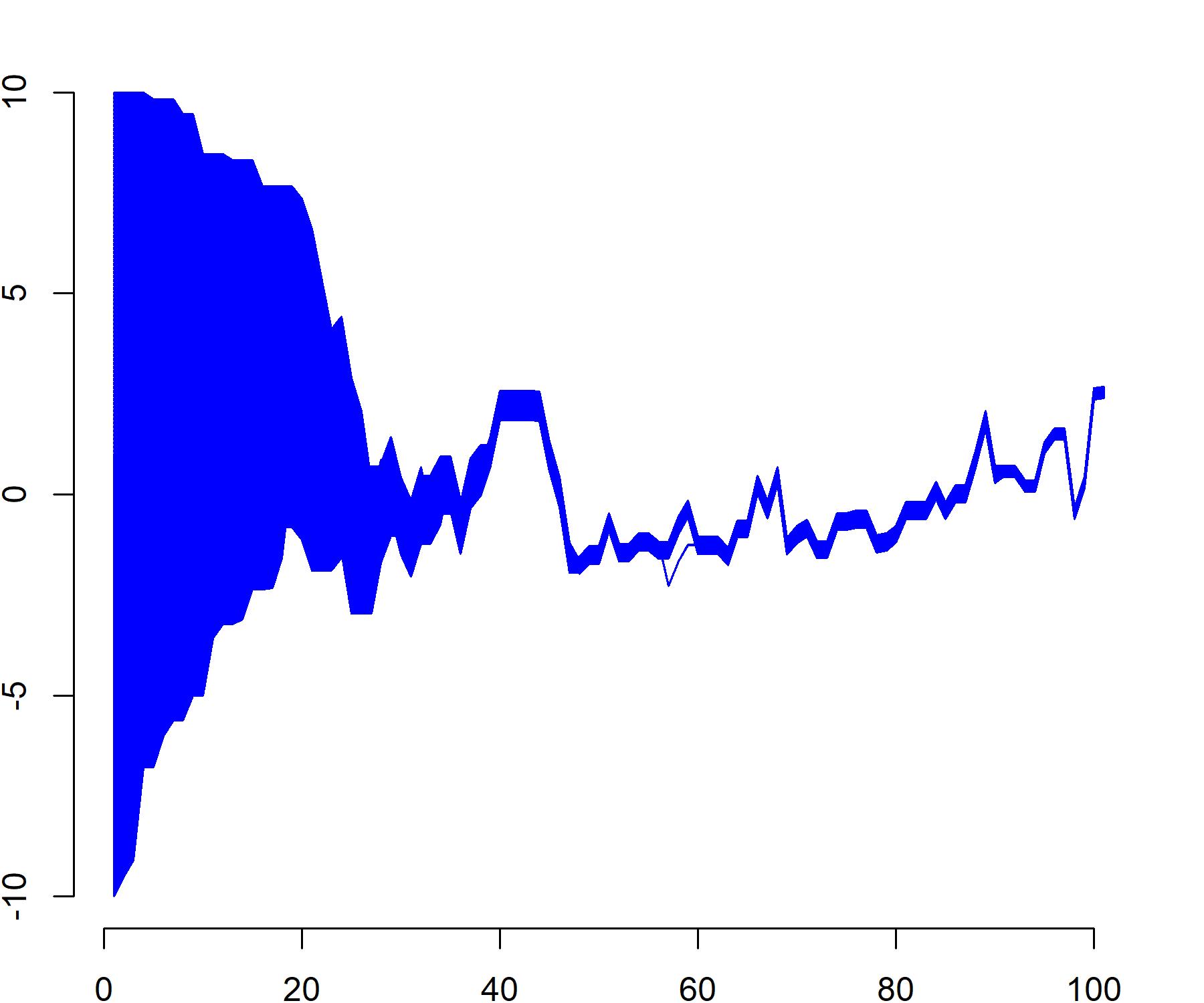}
\end{center}
\caption{Applying Metropolis simultaneously on $(-10,10)$ with target $N(0,1)$ and Markov chain $X_{t+1}= X_t + N(0,1)$. \label{fig:6}}
\end{figure}

The Metropolis algorithm stands as one of the most influential algorithms in Markov Chain Monte Carlo (MCMC) methods, recognized as one of the top ten algorithms of the 20th century \protect{\cite{dongarra}}. Let's run the traditional Metropolis algorithm simultaneously over the range $(-10,10)$ with the target distribution $N(0,1)$ and proposal $N(0,1)$. In each step, $X_{t+1}=X_t+Z_{t+1}$, where $Z_{t+1} \sim N(0,1)$. Figure \ref{fig:6} illustrates that all states within the considered the starting range move independently and converge towards the target distribution, albeit without coalescing. Despite the lack of monotonicity in the Metropolis algorithm, the starting range tends to converge and move within a narrow band.
To induce coalescence, one approach is to utilize the floor() function, as follows:
$$X_{t+1}=\lfloor X_t \rfloor + \Delta X_{t+1} \textnormal{  where  } \Delta X_{t+1} \sim  N(0,1).$$
Figure \ref{fig:7} illustrates the Metropolis algorithm with this new Markov chain, showcasing the occurrence of coalescence. It visually demonstrates the convergence of the state space within the interval $[n,n+1]$, where signals are emitted from each interval, leading to subsequent monotonic coalescence. 
By including floor function, the chain changed, and we cannot use Metropolis algorithm, since it needs a symmetric proposal, which leads us to normal multishift coupler.

\subsection{Metropolis-multishift coupler} \label{s1}

When applied simultaneously to the considered state space, the Metropolis algorithm fails to induce coalescence. A well-chosen proposal for the Metropolis sampler can significantly enhance convergence speed and induce coalescence. The symmetry of the normal multishift coupler makes it suitable for integration within the Metropolis algorithm framework. While this coupler does not require constructing a specific Markov chain to achieve the target distribution, it is not strictly monotone but exhibits behavior closely resembling monotonicity. This unique characteristic distinguishes it from traditional Metropolis algorithms, particularly in scenarios involving shifting between modes, where coalescence occurs more rapidly compared to conventional approaches. The Metropolis algorithm with a multishift proposal offers two distinct advantages: (1) it eliminates the need for constructing a specific Markov Chain, and (2) it induces coalescence. The algorithm for Metropolis with a multishift coupler is outlined as follows:

\begin{quote}
    Start with an arbitrary point $X_0$,\\
    Generate proposal $y$ by multishift coupler based on previous value, $y=\phi(X_t, U_{t+1})$,\\
    Compute $\alpha=min \left(1, \frac{\pi(y)}{\pi(X_t)} \right), $\\
    Generate $u \sim Uniform(0,1)$, \\
    If $u \le \alpha $ set $X_{t+1}=y$,\\
    Else set $X_{t+1}=X_t $,\\
    Do recursively from second step until $t=T$.
\end{quote}
If the starting range is chosen significantly distant from the target distribution, or if the target distribution is multimodal, coalescence may occur at an incorrect position, or the chain may become trapped within one of the modes. To explore these scenarios, we will apply the Metropolis-multishift coupler to six different target distributions, which include both unimodal and multimodal distributions. The considered target distributions are:

\begin{enumerate}
  \item $N(0,1)$
  \item $N(30,1)$
  \item $0.8N(-2,1)+0.2N(2,1)$
  \item $0.2N(-5,1)+0.2N(5,1)+0.6N(15,1)$
  \item $0.8Uniform(-100,100)+0.2Beta(50,50)$
  \item $0.9Uniform(-100,100) +0.1Beta(500,500)$
\end{enumerate}

The Metropolis-multishift coupler implemented simultaneously over the range (-10,10) for the first case of $N(0,1)$. Figure \ref{fig:8} depicts the initial partitioning of the starting range into intervals or sources by this coupler, emitting one or several signals from each source. Subsequently, these signals tend to undergo monotonic coalescence. This behavior contrasts with that of the Metropolis algorithm, which uniformly moves the entire state space and converges it into a narrow band.

The Metropolis-multishift coupler is also applied to the range $(-10,10)$ for the target distribution $N(30,1)$. However, this choice of starting range appears to be unsuitable for this target distribution. Figure \ref{fig:9} demonstrates how the algorithm gradually adjusts towards the target position. Selecting an inappropriate starting range can result in several issues, including slower coalescence and potential coalescence occurring at incorrect positions, particularly when the starting range is significantly distant from the target. Nevertheless, by persisting with the chain, it eventually converges to the target distribution.

The third scenario features a multimodal target distribution, with modes located at -2 and 2. These modes are close enough for the multishift normal standard coupler to transition between them. We apply the Metropolis-multishift coupler to the range $(-30,30)$ with the target distribution $0.8N(-2,1)+0.2N(2,1)$. The results are illustrated in Figure \ref{fig:10}. The chain effectively shifts between both modes while preserving the respective weights of each mode.

The fourth scenario involves three modes that are relatively distant from each other, with a gap of 10 between each pair of modes. Given this setup, it's worth considering whether the multishift normal standard coupler can seamlessly transition between these modes. Can we anticipate a behavior similar to the third scenario, where the coupler moves between modes effortlessly? Figure \ref{fig:11} illustrates the behavior of the Metropolis-multishift of normal standard coupler. It shows that the coupler moves away from the initial state space and becomes trapped within the three modes.

\begin{figure}[ht]
\begin{center}
\includegraphics[width=3in]{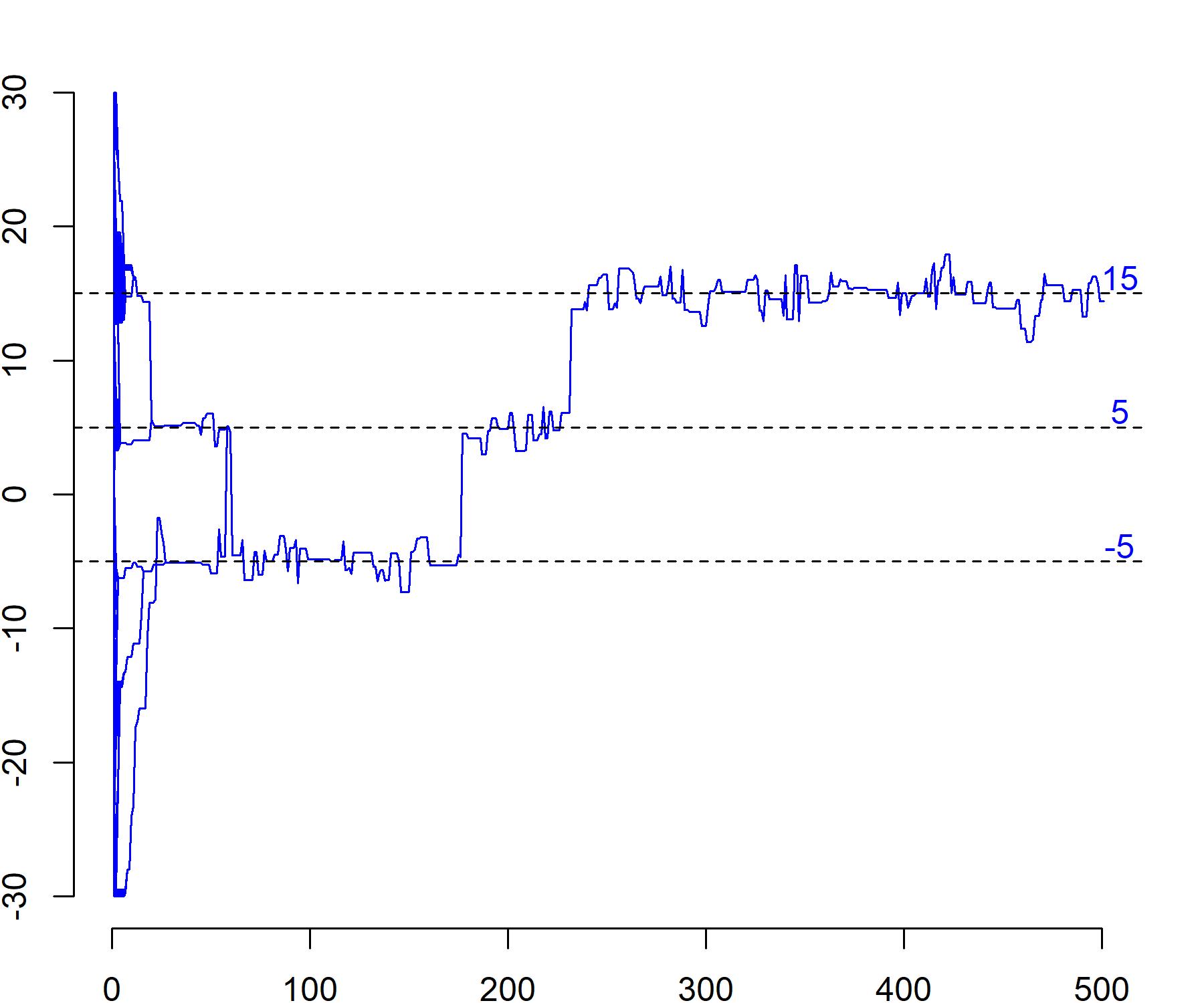}
\end{center}
\caption{Applying Metropolis-multishift simultaneously on $(-30,30)$ with target $0.2N(-5,1)+0.2N(5,1)+0.6N(15,1)$ and proposal multishift $N(0,3.5^2)$. \label{fig:12}}
\end{figure}

The gap between modes exceeds the deviation of the proposal distribution, resulting in the chains becoming trapped. In this case, we utilized the multishift of standard normal as the proposal for the Metropolis algorithm. To address this issue, a normal multishift with a larger scale, such as $N(0, 3.5^2)$, is needed. Choosing the appropriate scale significantly impacts the coalescence time. Figure \ref{fig:12} illustrates how the chain transitions between modes with the adjusted proposal distribution.

The fifth case comprises a target distribution with a mode around 0.4 and 0.6, featuring long tails spanning from -100 to 100, as depicted in Figure \ref{fig:13}, resembling a Dirac delta function overall, upon closer inspection, a beta density emerges. We apply the Metropolis-multishift coupler simultaneously over the range $(-100,100)$ with the target distribution $0.8Uniform(-100,100)+0.2Beta(50,50)$. Two different scales, $3.5^2$ and $1$, are considered for the multishift proposal. Given the wider range of this target distribution, coalescence occurs more rapidly with the larger scale, as illustrated in Figures \ref{fig:14} and \ref{fig:15}, where the mode is discernible in both cases.

In the sixth case, we increased the weight of the uniform distribution up to 0.9. While the mode is detectable in the figure, it becomes more ambiguous for larger weights. Despite the increased computational cost of the Metropolis-multishift method for targets such as the recent case, the method remains effective. It's important to note that not only are the weights changed, but the beta density is also altered to be narrower in the mode area. To enhance mode detection, the scale of the proposal is reduced, causing the chain to move more slowly. Consequently, the number of steps is increased to one and a half million. Figure \ref{fig:16} demonstrates that the chains coalesce after $500K$ steps, with the mode clearly detectable.

The Metropolis-multishift coupler is not monotone due to the acceptance-rejection process inherent in the Metropolis algorithm. However, it tends to exhibit nearly monotonic behavior towards the target distribution. Selecting two or more paths from the considered range does not alter the time of coalescence. Non-monotonic behavior typically arises in areas where the gap between paths exceeds the scale of the proposal. The size and proximity of the considered state space significantly affect the time of coalescence, particularly if it is large or distant from the target. Choosing an appropriate proposal, especially with a suitable scale, can decrease the time of coalescence. Therefore, the time of coalescence in the Metropolis-multishift coupler depends on: (1) the starting range and (2) the chosen scale of the proposal. The next two subsections delve into the convergence properties of the Metropolis-multishift coupler and the optimal ranges, with the goal of selecting suitable starting ranges and offering solutions for targets with unbounded state spaces.

\subsection{Convergence properties}
Let $\pi$ represent the target distribution. Consider a bounded starting range with a probability greater than $1-\epsilon$, where $\hat{0}$ and $\hat{1}$ denote the minimum and maximum of this range, respectively. The selection of the starting range and its impact on sampling is discussed in subsection \ref{sub1}. The chain $\{ \hat{0}_t \}$ starts from $\hat{0}$ at $t=0$, governed by the transition kernel of a Metropolis-multishift coupler with $\pi$ as its stationary distribution; the same applies to chain $\{ \hat{1}_t \}$. Initiate a primary chain $\{ X_t \}$ using the same Metropolis-multishift coupler from an arbitrary point $X_0$ at $t=0$ within the starting range, ensuring that all points in the starting range coalesce with the two extreme chains of $\hat{0}_t$ and $\hat{1}_t$ as auxiliary chains. The chains $\{ \hat{0}_t \}$, $\{\hat{1}_t \}$, and $\{X_t \}$ run simultaneously or jointly. Let $T$ be the length of a block such that $P(\text{coupling event at time } t \ | \ t < T) = p > 0$. The chain $\hat{0}_t$ updates to $\hat{0}$ when $mod(t,T)=0$, and similarly for chain $\hat{1}_t$.

 The coupling event inequality (\protect{\cite{lindvall1992lectures}}, \protect{\cite{thorisson2000appl}}) connects the distribution of coupling time to the convergence properties of an MCMC algorithm. We provide some definitions to explain this inequality for Metropolis-multishift coupler. 
 
 Let $X$ and $\hat{X}$ be random variables with distributions $\mathcal{L}(X)$ and $\mathcal{L}(\hat{X})$ respectively, defined on probability space $\left( \mathcal{X}, \mathcal{F} \right)$. The total variation between $\mathcal{L}(X)$ and $\mathcal{L}(\hat{X})$ is given by:
 $$\|\mathcal{L}(X)-\mathcal{L}(\hat{X})\|=2\sup_{A \in \mathcal{F}} \left( \mathcal{L}(X \in A)-\mathcal{L}(\hat{X} \in A)\right).$$
 
 Suppose $T_{\hat{0}}^*$ denotes the coupling time of Markov chains $\{X_t\}$, $\{\hat{0}_t\}$;  and $T_{\hat{1}}^*$ represents the coupling time of Markov chains $\{X_t\}$, $\{\hat{1}_t\}$ in a successful block with length $T$.
The coupling event inequality (\protect{\cite{lindvall1992lectures}}, \protect{\cite{thorisson2000appl}}) implies that
$$X_{T_{\hat{0}}^*}=\hat{0}_{T_{\hat{0}}^*} \Longrightarrow \|\mathcal{L}(X_t)-\mathcal{L}(\hat{0}_t)\| \le 2P(T_{\hat{0}}^* > t),$$
and similarly for $\hat{X}_{T_{\hat{1}}^*}=\hat{1}_{T_{\hat{1}}^*}$; define $T^*=max(T_{\hat{0}}^*,T_{\hat{1}}^*)$, hence 
\begin{equation} \label{eq1}
\|\mathcal{L}(X_t)-\mathcal{L}(\hat{0}_t)\|+\|\mathcal{L}(X_t)-\mathcal{L}(\hat{1}_t)\|  \le 4P(T^* > t),
\end{equation}
where $T^*$ denotes the coupling time of Markov chains $\{X_t\}$, $\{\hat{0}_t\}$, and $\{\hat{1}_t\}$. 

The ergodic properties of the Metropolis algorithm have been thoroughly investigated by \protect{\cite{andrieu2006ergodicity}}, \protect{\cite{atchade2007geometric}}, and \protect{\cite{wang2017geometric}}, particularly regarding symmetric proposals. Let's investigate this property for the Metropolis-multishift coupler. Consider $P$ as a $\pi$-irreducible Metropolis-multishift kernel, with a one-step transition kernel denoted as $P(x,dy)$ and a $t$-step transition kernel as $P^t(x,dy)$, where $\pi$ represents the stationary distribution. We denote the joint density of two independent draws from $\pi$ as $\pi^2$, and $Q^t$ represents the joint distribution of $t$ iterates in the primary Markov chain ${X_t}$:
 $$Q^t(dx,dy)=P^t(x,dy)\pi(dx).$$
 Let the auxiliary chain $\hat{0}_t$ starts from $\hat{0}$, in a successive block $t < T$; then $Q_{\hat{0}}^t(dy)=P^t(\hat{0},dy)$ and the same for $\hat{1}_t$. By using Theorem 1 of \protect{\cite{johnson1998coupling}}
\begin{equation} \label{eq2}
  \|\pi^2-Q^t\| \le 2\|Q_{\hat{0}}^t-Q^t\| \quad \textnormal{and} \quad \|\pi^2-Q^t\| \le 2\|Q_{\hat{1}}^t-Q^t\|,
\end{equation}
hence by using (\ref{eq1})
\begin{equation} \label{eq3}
  \|\pi^2-Q^t\| \le 4P(T^* > t),
\end{equation}
where $P(T^* > t)$ is the proportion of times that the Markov chains $\{X_t\}$, $\{\hat{0}_t\}$, and $\{\hat{1}_t\}$ fail to couple in $t$ or fewer steps. Note that if the number of auxiliary paths increases from 2 paths to $k$ paths, inequality \ref{eq3} still holds, where $T^*$ represents the coupling time of $k$ auxiliary paths and the primary chain ${X_t}$. In fact, if the Metropolis-multishift coupler coalesces all the starting range to a single point, then the convergence remains independent of the number of auxiliary paths. Subsection \ref{sub2} demonstrates that the two extreme auxiliary paths are sufficient for the Metropolis-multishift coupler to coalesce the entire starting range in $t$ steps.

\begin{Prop}
    Suppose $Q^t$ represents the joint distribution of $t$ iterations of the Markov chain $\{ X_t \}$ governed by a $\pi$-irreducible Metropolis-multishift coupler, and $T^*$ is the coupling time of the two extreme auxiliary chains $\{ \hat{0}_t \}$ and $\{ \hat{1}_t \}$ with the primary chain $\{ X_t \}$. In this context, inequality \ref{eq3} holds.
\end{Prop}

Proposition 1 offers an upper bound for total variation, and if it decreases rapidly with an increase in $t$, then inequality \ref{eq3} serves as a mechanism for generating samples using the Metropolis-multishift coupler. The Metropolis-multishift coupler is uniformly ergodic if there exist positive constants $M$ and $r < 1$ such that $\|\pi^2-Q^t\| \le Mr^t$, \protect{\cite{tierney1994markov}}. In the remainder of this subsection, the goal is to explain $P(T^* > t)$, supported by a simulation to provide justification.
 
Suppose a Metropolis-multishift coupler operates within a suitable starting range simultaneously, as discussed in Subsection \ref{sub1}. The multishift coupler, acting as the proposal, is a step function that partitions the starting range into intervals and proposes a unique value for each interval. 

In particular, the mapping $\lfloor \frac{s+R-X}{R-L} \rfloor (R-L) + X$ proposes a unique value for all $s \in [X-L,X+R]$. It's worth noting that $L$ and $R$ are randomly generated from a normal distribution at each step, while $X$ is uniformly selected from the interval $[L,R]$. Once $X$ is generated, the algorithm creates an interval around it. Similar intervals are then constructed on the left and right sides of this interval, effectively dividing the starting range into intervals. The mapping then proposes a unique value for each of these intervals.

The proposed value is accepted with a probability of $\alpha=min\{1, \frac{\pi(X_t)}{\pi(X_{t-1})}\}$ and rejected with a probability of $(1-\frac{\pi(X_t)}{\pi(X_{t-1})})$. It's important to note that when the proposed value is rejected, $\frac{\pi(X_t)}{\pi(X_{t-1})} < 1$. 

Let's consider a sufficiently large $t$ for the coalescence of all paths. Since the starting range is bounded, achieving the coalescence of all paths requires at least a minimum number of successive iterations, let's say $m$ iterates. Then, the probability of non-coalescence implies that the number of successive iterations should be less than $m$. Define $M$ as the maximum of all $m-1$  values of $\alpha$, and $r$ as the maximum of rejected probabilities of $(1-\frac{\pi(X_t)}{\pi(X_{t-1})})$. Note that $\pi(X_t) > 0$, as if $\pi(X_t)=0$, then $X_t$ would be within the null set.  Hence,
\begin{equation} \label{eq4}
    P(T^*>t) \leq M^{m-1} r^{t-m+1},
\end{equation}

where $m < t$, and $0<r<1$. This heuristic explanation demonstrates that $P(T^* > t)$ decays exponentially as the number of iterations $t$ increases. 

\begin{figure}[ht]
\begin{center}
\includegraphics[width=3.5in]{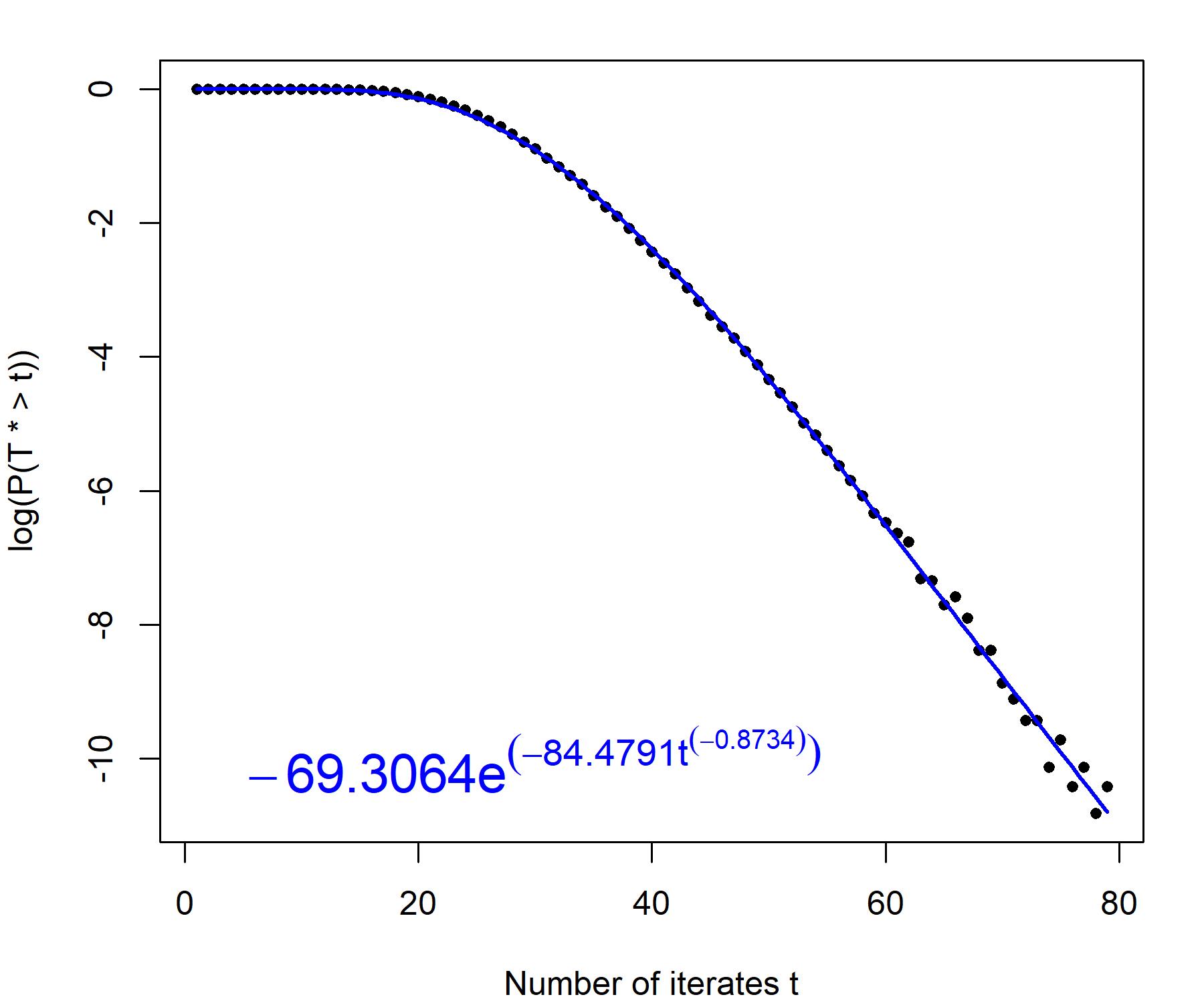}
\end{center}
\caption{Logarithms of $P(T^*>t)$ versus $t$ with 100k replications for each t. \label{fig:21}}
\end{figure}

In order to demonstrate the exponential decrease of $P(T^* > t)$, we executed the Metropolis-multishift coupler on three paths starting from (-10, 0, 10) for a target of $N(0,1)$ with $100K$ replications for each $t=1$ to $100$. Figure \ref{fig:21} depicts the results for $t=1$ to $79$, as beyond $t=79$, $P(T^* > t)$ becomes zero. This indicates that replications more than $100K$ are required to observe at least one run without coalescence for $t=80$ and beyond, given the exceedingly small probability of non-coalescence.

It's worth noting that $P(T^* > t) = 1$ for the first 7 observations, signifying that the Metropolis-multishift coupler requires a minimum of 8 successive steps for the first coalescence, as described by $m$ in equation \ref{eq4}. The fitted curve for $log(P(T^* > t))$ from $t=1$ to $79$ is $log(P(T^* > t)) = -69.3064e^{-84.4791t^{-0.8734}}$. Extrapolating for $t=150$, the fit yields $P(T^* > t) = e^{-23.96207}$.

According to Theorem 2 of \protect{\cite{johnson1998coupling}}, for $n$ independent samples from target $\pi(\cdot)$, the total variation is bounded by $\|\pi^n-Q^t_n\| \le 4nP(T^* > t)$. If we take a sample size of $n=1000K$ and a block length of $T=150$, then the total variation from the target is bounded by $\|\pi^n-Q^t_n\| \le 4000000e^{-23.96207} \le 0.000157$. Figure \ref{fig:21} illustrates an exponential decay much stronger than the linear decays depicted in \protect{\cite{johnson1998coupling}}.

\subsection{The most interest range} \label{sub1}

\begin{figure}[ht]
\begin{center}
\includegraphics[width=4.5in]{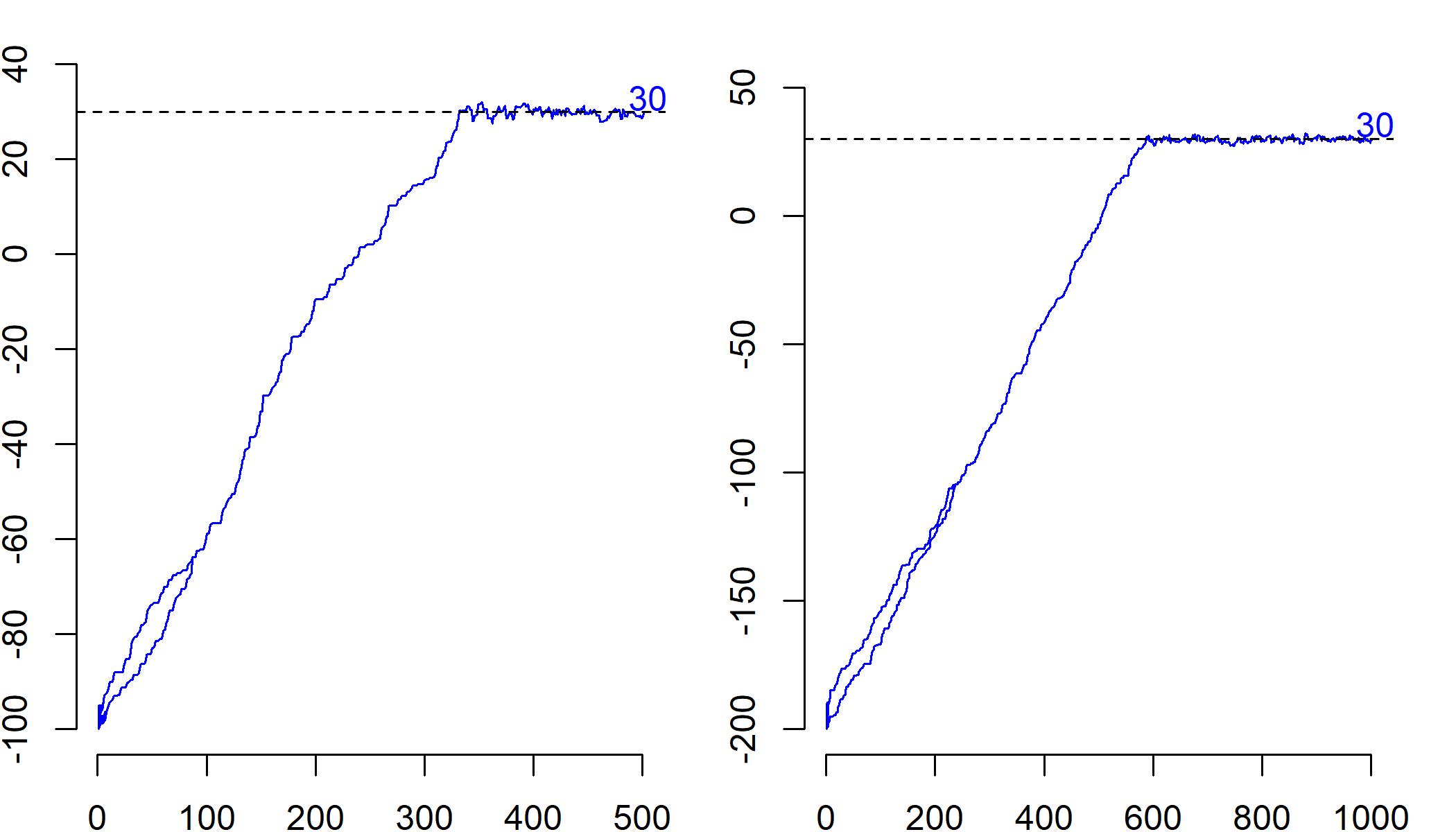}
\end{center}
\caption{An illustration of Metropolis-multishift applied simultaneously to the target distribution $(-30,30)$ with a mean of $N(30,1)$ and a proposal distribution of $N(0,1)$, starting from different ranges. \label{fig:18}}
\end{figure}

In the Subsection \ref{s1}, we discussed two factors influencing fast coalescence: the starting range and the scale of the proposal. Here, we will delve into the starting range. Selecting a starting range significantly distant from the target range may cause the two starting chains to coalesce at a point far from the target distribution. When combined with a small value of $T$, this can introduce bias into the sampling process. For instance, in Figure \ref{fig:18}, where the target in both cases is $N(30,1)$; the first case selects a start range of $[-100,-95]$, resulting in coalescence around -70 after 100 steps. Although, in the second case, the start range is $[-200,-190]$, leading to coalescence around -100 after 200 steps!

When dealing with mixture distributions featuring multiple modes, selecting the wrong starting range alongside a small scale can lead to chains becoming trapped in a specific mode or prevent coalescence altogether. For instance, in the fourth case with three modes and a small-scale proposal, Figure \ref{fig:17} illustrates how adjusting the range can inadvertently trap chains in one mode, potentially misleading experimenters into believing the algorithm is functioning correctly. Conversely, opting for a larger starting range with a small scale may result in chains getting stuck in two modes, completely avoiding coalescence.

In another scenario where three modes have varying gaps, selecting a larger scale and an incorrect starting range can result in rapid coalescence with a shorter block length $T$, leading to biased samples that favor the two closer modes. Such issues arise from making the wrong choice of range. Evaluating the target distribution $\pi(x)$ by graphing it can assist in selecting an appropriate starting range. While this process may not be straightforward, it provides a general understanding of the distribution's modes, aiding in better range selection.

\protect{\cite{foss1998perfect}} highlighted the necessity of a uniformly ergodic underlying Markov chain for the proper functioning of CFTP. However, numerous common chains utilized in Bayesian inference possess unbounded state spaces and lack uniform ergodicity. \protect{\cite{murdoch2000exact}} tackled this challenge by proposing solutions through various Markov chain constructions. Additionally, \protect{\cite{wilson2000layered}} introduced $\hat{0}$ and $\hat{1}$ for bounded state spaces. Here, we delve into a more practical solution to address this issue.

When considering the standard normal distribution, most normal tables provide a largest bound of $[-3.59,3.59]$, with $P([-3.59,3.59])=0.9996$. Therefore, selecting the range $[-10,10]$ for the target $N(0,1)$ ensures that almost all samples fall within this range, which is more than sufficient for practical purposes. The likelihood of generating a sample outside the range $[-10,10]$ is practically zero. Consequently, incurring significant computational costs for such a negligible probability is unreasonable, leading us to define the "Most Interesting Range" (MIR).

\begin{Def}
Consider a Markov chain with target $\pi$, state space $\mathcal{X}$, and the interested probability of $1-\varepsilon$, sufficiently large, and define $\mathbb{F}_\varepsilon=\{ A \subset \mathcal{X}: \pi(A) \ge 1-\varepsilon \}$. Let $\mu$ be Lebesgue measure for continuous state spaces or cardinality for discrete state spaces, then
$$MIR=\arginf_{A \in \mathbb{F}_\varepsilon} \mu(A).$$
\end{Def}

MIR offers two significant advantages: (1) it provides a small, bounded starting range, and (2) it selects the correct part of the state space, steering clear of positions distant from the target. By this definition, we can establish $\hat{0}=\inf(MIR)$ and $\hat{1}=\sup(MIR)$, with $MIR^C=\mathcal{X} \setminus MIR$ representing the unlikely range where $\pi(MIR^C)=\varepsilon$. The $MIR$ encourages the selection of a range around the mode(s). While identifying the $MIR$ in complex posteriors is challenging, keeping it in mind proves invaluable.

Now, the question arises: Can the algorithm generate a sample from the unlikely range? If so, what is the probability?

Suppose $\phi: \mathcal{X} \longrightarrow \mathcal{X}$ represents the random operation of the Metropolis-multishift, running with two auxiliary chains $\{ \hat{0}_t \}$ and $\{ \hat{1}_t \}$, and a primary chain $\{ X_t \}$ starting from $\hat{0}$, $\hat{1}$, and an arbitrary position $X_0$ on the MIR, respectively. In a successive block, the coalescence of the three chains $\hat{0}_t$, $\hat{1}_t$, $X_t$ indicates that the primary chain has forgotten its past states. In fact, $X_{T} = \phi_{T} \circ \phi_{T-1} \circ \cdots \circ \phi_{0}(\hat{0})= \phi_{T} \circ \phi_{T-1} \circ \cdots \circ \phi_{0}(\hat{1})$ does not depend on $\hat{0}$, $\hat{1}$ or any $x \in MIR$. If $T$ is sufficiently large, then $X_T$ converges to the stationary distribution $\pi$, as indicated by $\int_{A} P(\phi(x) \in A) \pi(dx) = \pi(A)$, and becomes independent of $X_0$. Note that the pair $(C, \ X_T)$ solely depends on randomness $\{ U_t\}_{t=1}^{t=T}$. In the case of such a sufficiently large $T$: 
$$P\left( X_0 \in MIR, \ X_T \in MIR^C \ | \ C  \right)=(1-\epsilon)\epsilon.$$
where $C$ is the coupling event of the three chains $\hat{0}_t=\hat{1}_t=X_t$ and $t < T$. In ROCFTP, the diagnosis mechanism ensures convergence and independent draws from the target distribution $\pi$. Therefore, in ROCFTP with the Metropolis-multishift coupler, it is sufficient to choose the length of the block $T$ such that $P\left( C \ | \ T\right)=p > 0$.

\begin{Prop}
    Choosing $MIR$ as the starting range reduces the likelihood of the Metropolis-multishift coupler reaching $MIR^C$ by a factor of $1 - \epsilon$.
\end{Prop}

The probability that a chain attains a value from the unlikely range is $(1-\varepsilon)\varepsilon$. So, selecting MIR effects ROCFTP hitting $MIR^C$ by factor of $(1-\varepsilon)$ which is proper for practical purpose.

\subsection{How many starting paths are needed?} \label{sub2}
To reduce the computational cost of ROCFTP, one strategy is to minimize the number of starting paths. When dealing with a monotone coupler and a bounded state space, it suffices to demonstrate the coalescence of two paths originating from $\hat{0}$ and $\hat{1}$. The monotonic property of a coupler ensures that all points in the state space will ultimately coalesce. The Metropolis-multishift coupler is not strictly monotone due to its acceptance/rejection process, it exhibits behavior close to monotonicity, largely thanks to the multishift coupler.

If the state space is continuous, the multishift coupler transforms the entire state space into a countable set of randomly distributed points. This characteristic underscores the effectiveness of the multishift coupler, as these points progressively coalesce in each subsequent step. Figure \ref{fig:19} visually demonstrates how the Metropolis-multishift coupler condenses the interval $[ -2, 2]$ within four steps to three points.

We aim to demonstrate that the near-monotone behavior of the Metropolis-multishift coupler is sufficient for use in ROCFTP. This implies that if paths $\hat{0}_t$ and $\hat{1}_t$ coalesce, then the entire state space will also coalesce. Therefore, constructing ROCFTP based on the two paths $\hat{0}_t$ and $\hat{1}_t$ is adequate.

To provide a simulation justification, we considered the first four targets introduced in Subsection \ref{s1}, representing the most critical cases. We calculated the time of coalescence simultaneously for 2, 10, 100, 1000, and 10000 paths. Simultaneously, in this context, means that these paths are inclusive of each other; for instance, the experiment with 100 paths encompasses the experiment with 10 paths. In fact, these experiments run jointly using the same randomness.

Table \ref{T2} displays the mean time of coalescence for the selected four targets, calculated from 1000 replications. It shows minimal differences in the mean coalescence time between two paths and more paths, especially in the first three targets. This indicates that whether we start with two paths or 10000 paths, the results are nearly identical, with all paths coalescing at almost the same time. In essence, any increase in coalescence time due to using more paths is negligible. The results highlights that if the two paths $\hat{0}_t$ and $\hat{1}_t$ coalesce, then the entire state space will also coalesce, even if it takes slightly longer.
\begin{table}[ht]
\footnotesize
\begin{center}
\begin{tabular}{lccccc}
\hline \hline
Target                          & MIR         & Proposal & 2 Paths    & 10 Paths    & 100 Paths \\ \hline

$N(0,1)$                          & $[-10,10]$ & $N(0,1)$   & 29.663     & 29.671      & 29.678    \\
$N(30,1)$                         & $[20,40]$  & $N(0,1)$   & 29.892     & 29.901      & 29.901    \\
$0.8N(-2,1)+0.2N(2,1)$            & $[-10,10]$ & $N(0,1)$   & 42.542     & 42.546      & 42.546    \\
$0.2N(-5,1)+0.2N(5,1)+0.6N(15,1)$ & $[-15,25]$ & $N(0,3.5^2)$ & 148.320    & 150.978     & 151.343   \\
                                &              &          &            &             &           \\
Target                          & MIR         & Proposal & 1000 Paths & 10000 Paths &           \\ \hline

$N(0,1)$                          & $[-10,10]$ & $N(0,1)$   & 29.678     & 29.678      &           \\
$N(30,1)$                         & $[20,40]$  & $N(0,1)$   & 29.901     & 29.901      &           \\
$0.8N(-2,1)+0.2N(2,1)$            & $[-10,10]$ & $N(0,1)$   & 42.546     & 42.546      &           \\
$0.2N(-5,1)+0.2N(5,1)+0.6N(15,1)$ & $[-15,25]$ & $N(0,3.5^2)$ & 151.343    & 151.343     &           \\ \hline \hline
\end{tabular}
\end{center}
\caption{Simulating the mean coalescence time of the Metropolis-multishift coupler with different targets based on 1000 independent iterations. \label{T2}}
\end{table}

Table \ref{T3} presents the percentage of equal coalescence times with different numbers of starting paths. In most cases, whether the chains start from $\hat{0}_t$ and $\hat{1}_t$ or from the entire state space, they coalesce at the same time. The number of starting paths does not significantly affect the coalescence time. This observation underscores the nearly monotone behavior of the Metropolis-multishift coupler.

\begin{table}[ht]
\footnotesize
\begin{center}
\begin{tabular}{lccccc}
\hline \hline
Target                          & 2 Paths & 10 Paths & 100 Paths & 1000 Paths & 10000 Paths \\ \hline
$N(0,1)$                          & 99.2    & 99.9     & 100       & 100        & 100         \\
$N(30,1)$                         & 99.6    & 100      & 100       & 100        & 100         \\
$0.8N(-2,1)+0.2N(2,1)$            & 99.9    & 100      & 100       & 100        & 100         \\
$0.2N(-5,1)+0.2N(5,1)+0.6N(15,1)$ & 97.6    & 99.6     & 100       & 100        & 100         \\ \hline \hline
\end{tabular}
\end{center}
\caption{The percentage of equal coalescence times with different numbers of starting paths for the Metropolis-multishift coupler. The range and proposal settings are consistent with those in Table \ref{T2}. \label{T3}}
\end{table}

Table \ref{T4} presents the summary statistics of coalescence time with two paths, based on 10000 replications. It shows a notable gap between the central tendency and maximum values, indicating a long tail on the right side of the distribution of coalescence time (refer to Figure \ref{fig:20}). Based on the results in this subsection, we will construct ROCFTP using the Metropolis-multishift coupler with paths starting from $\hat{0}$ and $\hat{1}$.

\begin{table}[ht]
\footnotesize
\begin{center}
 \begin{tabular}{lcccccl}
\hline \hline
Target                          & Min  & 1st Q. & Median & Mean  & 3rd Q. & Max     \\ \hline
N(0,1)                          & 7.00 & 24.00  & 29.00  & 29.59 & 34.00  & 84.00   \\
N(30,1)                         & 7.00 & 24.00  & 29.00  & 29.60 & 34.00  & 90.00   \\
0.8N(-2,1)+0.2N(2,1)            & 9.00 & 29.00  & 38.00  & 42.59 & 51.00  & 259.00  \\
0.2N(-5,1)+0.2N(5,1)+0.6N(15,1) & 3.00 & 62.00  & 116.00 & 151.1 & 202.00 & 1384.00 \\ \hline \hline
\end{tabular}
\end{center}
\caption{Summary statistics of coalescence time using the Metropolis-multishift coupler with different targets and two starting paths. The range and proposal settings are consistent with those in Table \ref{T2}. \label{T4}}
\end{table}

\clearpage
\section{ROCFTP with Metropolis-multishift coupler}
ROCFTP initiates sampling from the second coupling event, where the first coupling event establishes the primary chain for sampling. The two auxiliary chains $\{ \hat{0}_t \}$ and $\{ \hat{1}_t \}$ function as a diagnostic mechanism to select an independent sample from the stationary distribution. The convergence of the chain to the stationary distribution $\pi$ is established by the first coupling event.

We are prepared to sample using ROCFTP with the Metropolis-multishift coupler. In this context, we will conduct ROCFTP with the Metropolis-multishift coupler on the four targets outlined in Table \ref{T2}, along with their related MIRs and proposals. The median time of coalescence from Table \ref{T4} will be used as the block length $T$. To assess goodness of fit, we will use Q-Q plots based on $10K$ samples from the targets. Figure \ref{fig:22} illustrates a comparison between the histogram and the related theoretical density, and includes the Q-Q plot for each target.

In panel (d) of Figure \ref{fig:22}, the histogram aligns well with the theoretical density, but the QQ plot reveals gaps among the modes. These outliers in the QQ plot appear unusual because they deviate from the expected proportions. Ideally, there should be around 2000, 2000, and 6000 samples around each mode according to the target distribution. However, ROCFTP with the Metropolis-multishift coupler may generate samples that don't exactly match these proportions. As a result, if the generated samples are more or less than 2000 for the first left mode, the QQ plot will show outliers either above or below the QQ line.

To identify these outliers, we calculate the absolute difference between the practical and theoretical quantiles, as shown in Figure \ref{fig:23}. There are 30 outliers detected, which amounts to about 0.3\% of all the samples. Despite these outliers, ROCFTP with the Metropolis-multishift coupler achieves a close match to the expected proportions overall.

\section{Discussion}
ROCFTP with the Metropolis-multishift coupler provides perfect samples for complex posteriors and converges rapidly to the target distribution at an exponential rate. It achieves faster convergence compared to the algorithm proposed by \protect{\cite{johnson1998coupling}}. Moreover, it is applicable to posteriors with unbounded state spaces by leveraging the $MIR$, which influences the algorithm to hit $MIR^C$ by a factor of $1-\epsilon$. Additionally, its performance on mixture distributions has been investigated.

\protect{\cite{berthelsen2010perfect}} utilized ROCFTP with the Gibbs sampler to generate samples for mixture models, but they did not examine the performance and properties of ROCFTP with the Gibbs sampler. On the other hand, the approach by \protect{\cite{jacob2020unbiased}} provides an unbiased estimate at a lower computational cost. However, ROCFTP with the Metropolis-multishift coupler provides exact samples, making it suitable for estimating and investigating the properties of more complex parameters in posteriors.

\newpage
\bigskip
\bibliographystyle{agsm}
\bibliography{Bibliography-MM-MC}

\clearpage
\bigskip
\appendix
\counterwithin{figure}{section}
\section{Appendix}
\subsection{Figures}

\begin{figure}[ht]
\begin{center}
\includegraphics[width=4.5in]{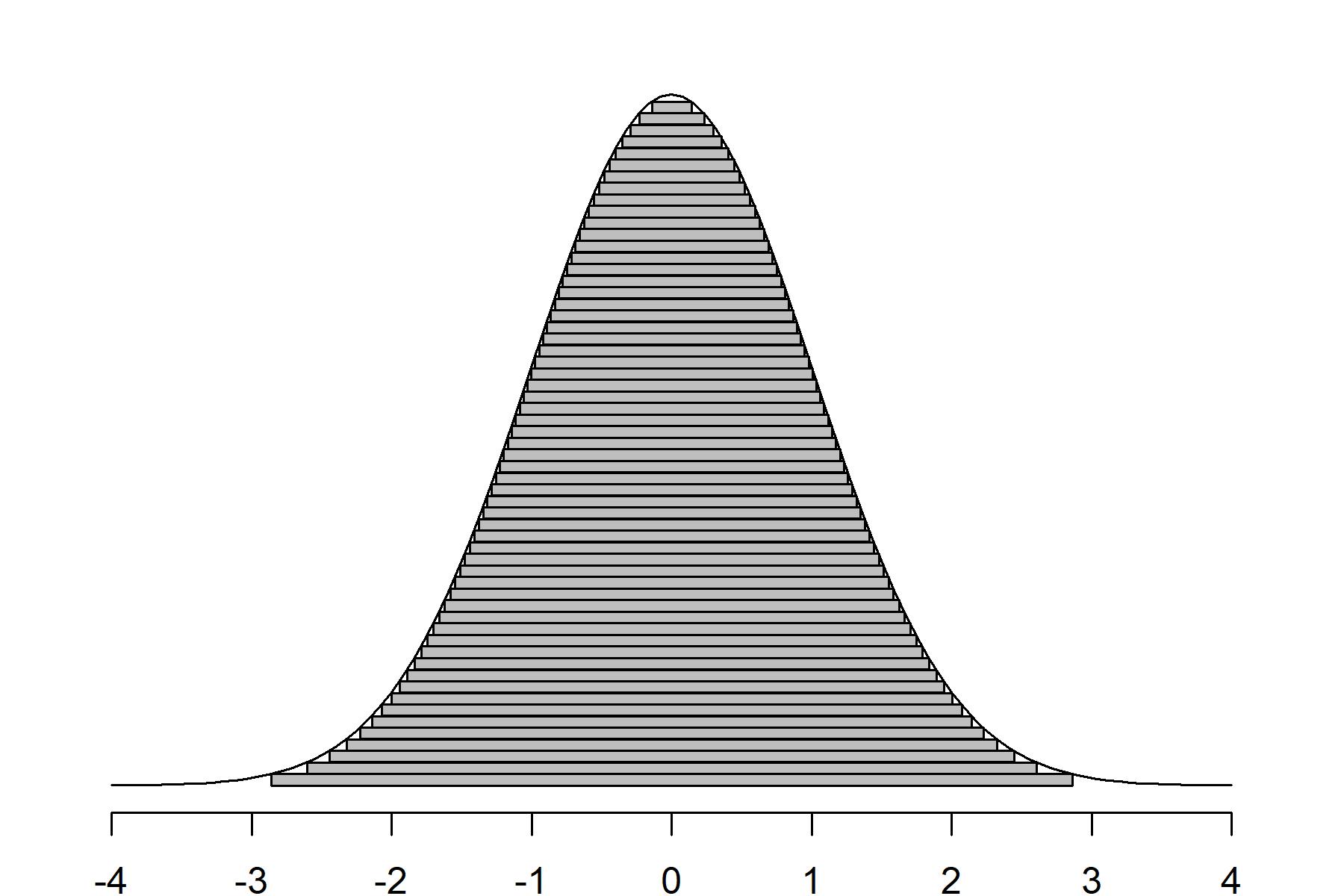}
\end{center}
\caption{Expression of normal distribution as a combination of layered uniform rectangles. \label{fig:1}}
\end{figure}

\begin{figure}[ht]
\begin{center}
\includegraphics[width=3in]{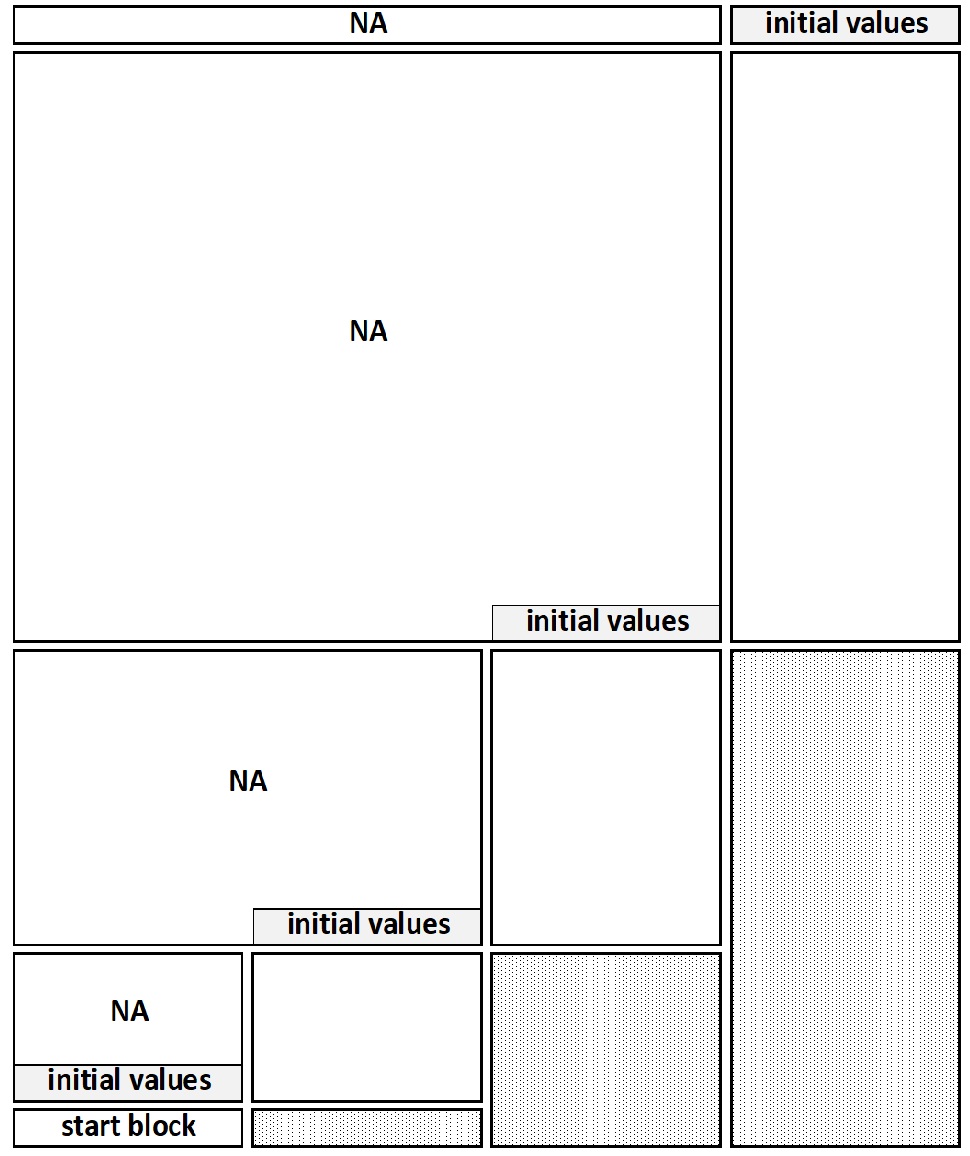}
\end{center}
\caption{Storing scheme of randomness in a matrix for CFTP algorithm. The storing process starts from the left, and the white and dotted blocks under initial values determine the new and previous randomness in each step. \label{fig:4}}
\end{figure}

\begin{figure}[ht]
\begin{center}
\includegraphics[width=4in]{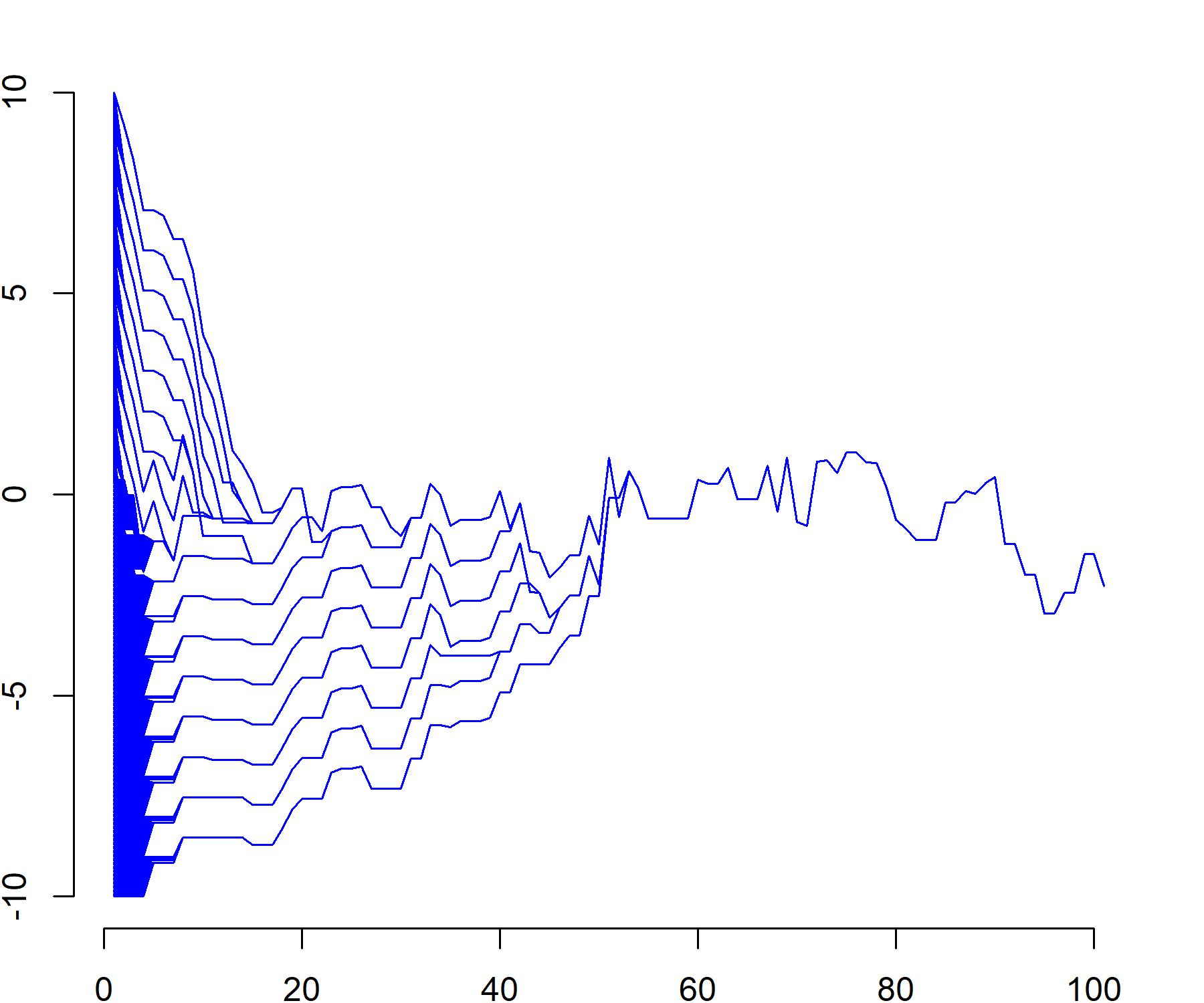}
\end{center}
\caption{Applying Metropolis simultaneously on $(-10,10)$ with target $N(0,1)$ and Markov chain $X_{t+1}=\lfloor X_t \rfloor + N(0,1)$. \label{fig:7}}
\end{figure}

\begin{figure}[ht]
\begin{center}
\includegraphics[width=3in]{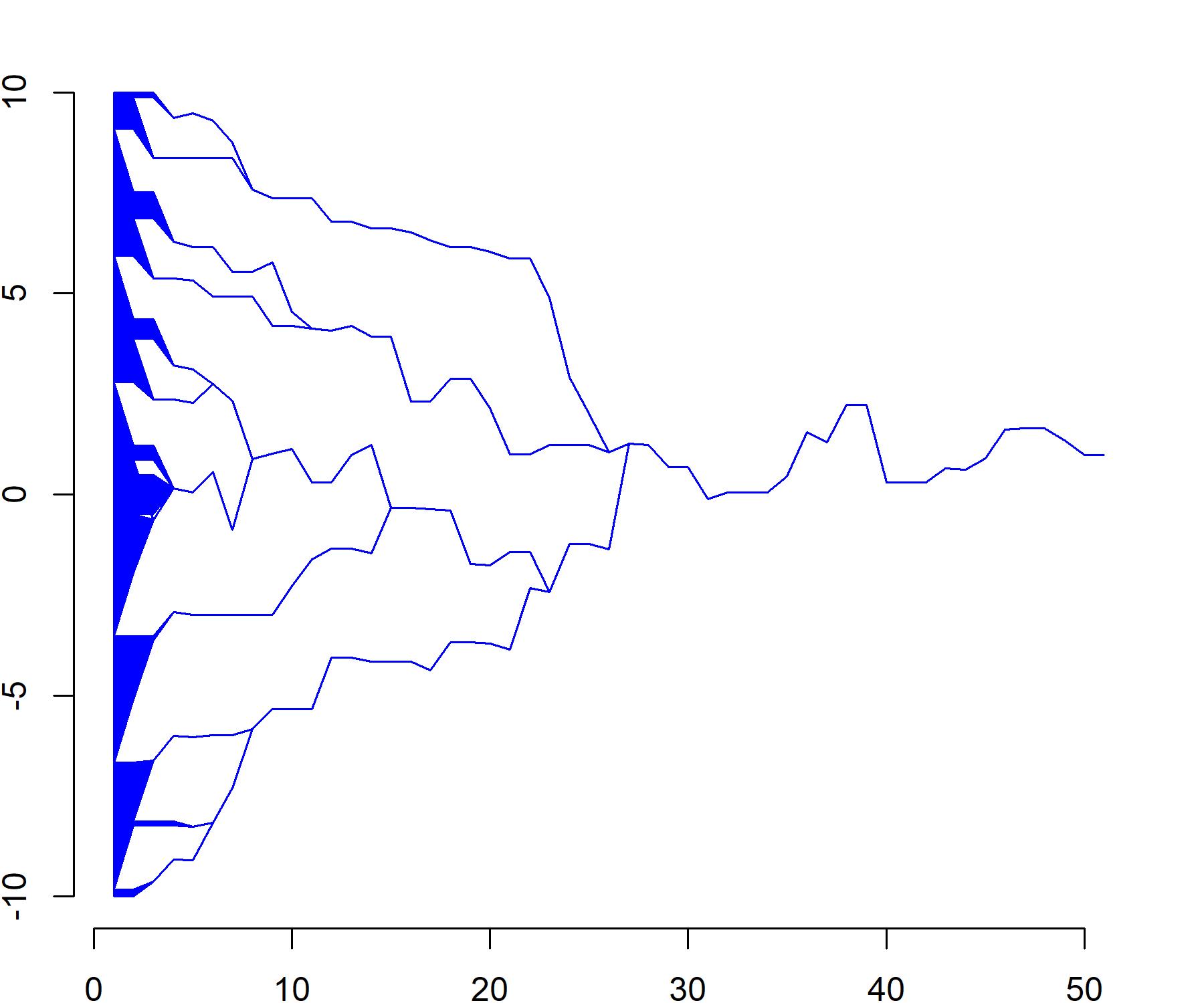}
\end{center}
\caption{Applying Metropolis-multishift simultaneously on $(-10,10)$ with target $N(0,1)$. \label{fig:8}}
\end{figure}

\begin{figure}[ht]
\begin{center}
\includegraphics[width=3in]{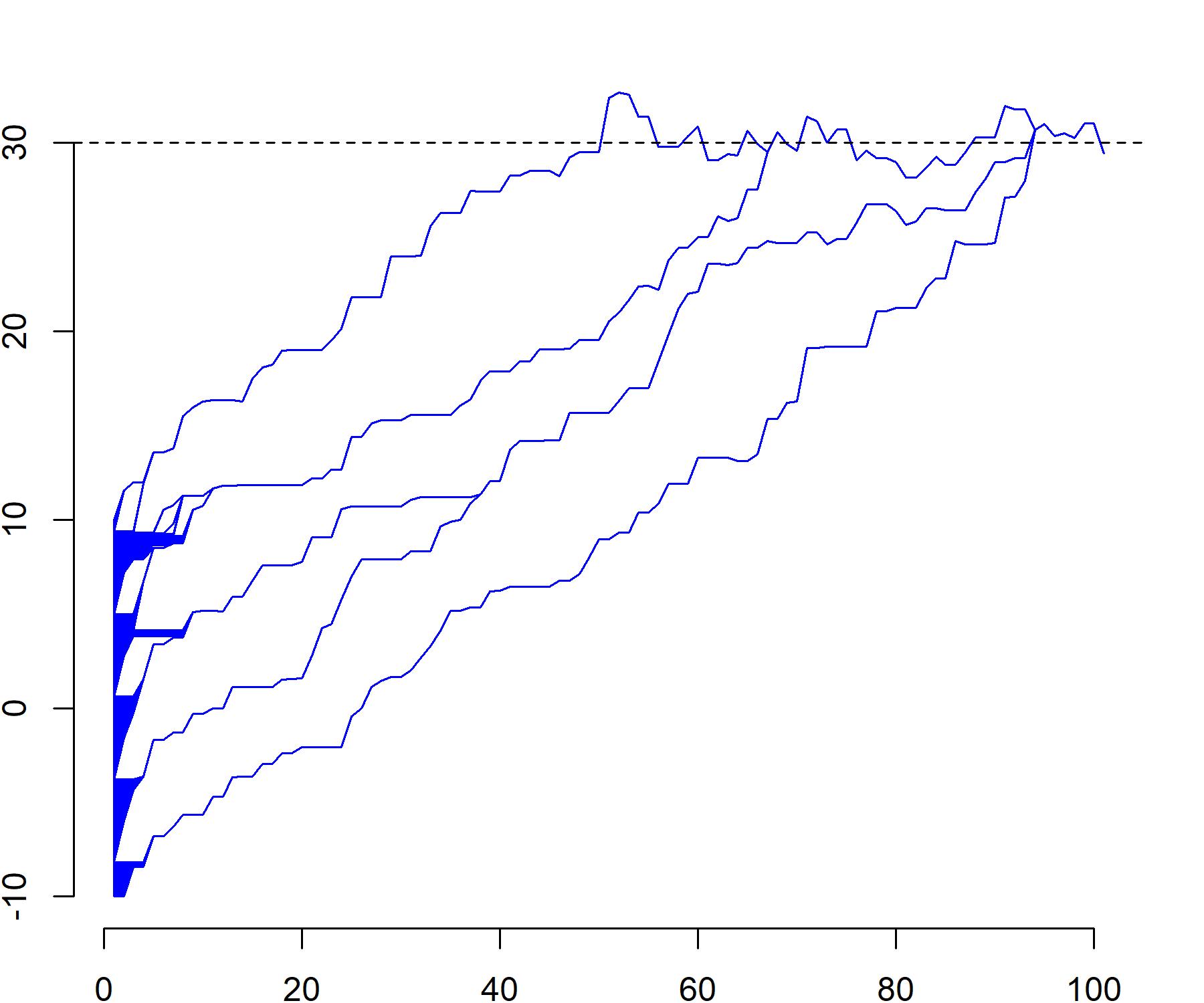}
\end{center}
\caption{Applying Metropolis-multishift simultaneously on $(-10,10)$ with target $N(30,1)$. \label{fig:9}}
\end{figure}

\begin{figure}[ht]
\begin{center}
\includegraphics[width=3in]{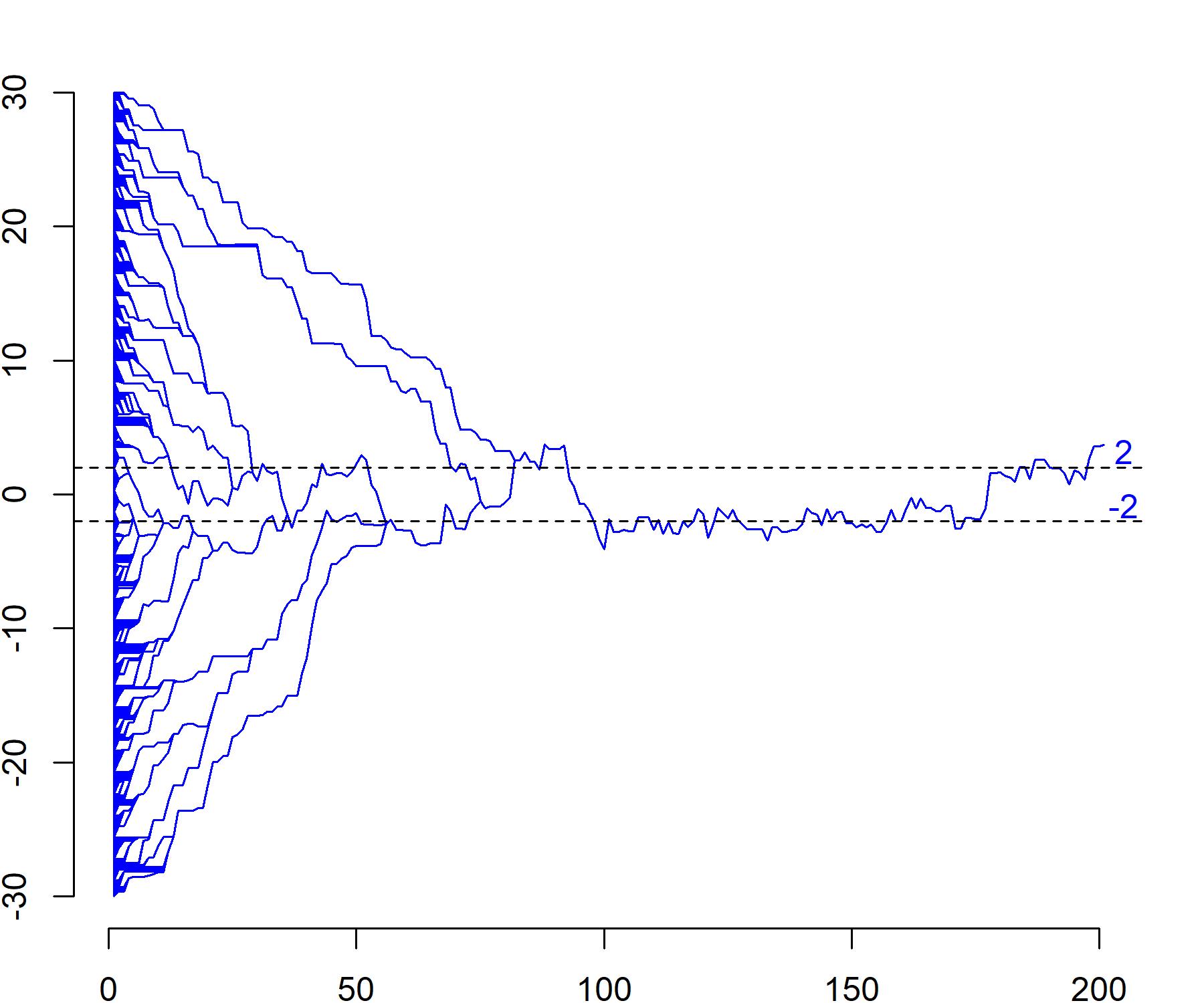}
\end{center}
\caption{Applying Metropolis-multishift simultaneously on $(-30,30)$ with target $0.8N(-2,1)+0.2N(2,1)$. \label{fig:10}}
\end{figure}

\begin{figure}[ht]
\begin{center}
\includegraphics[width=3in]{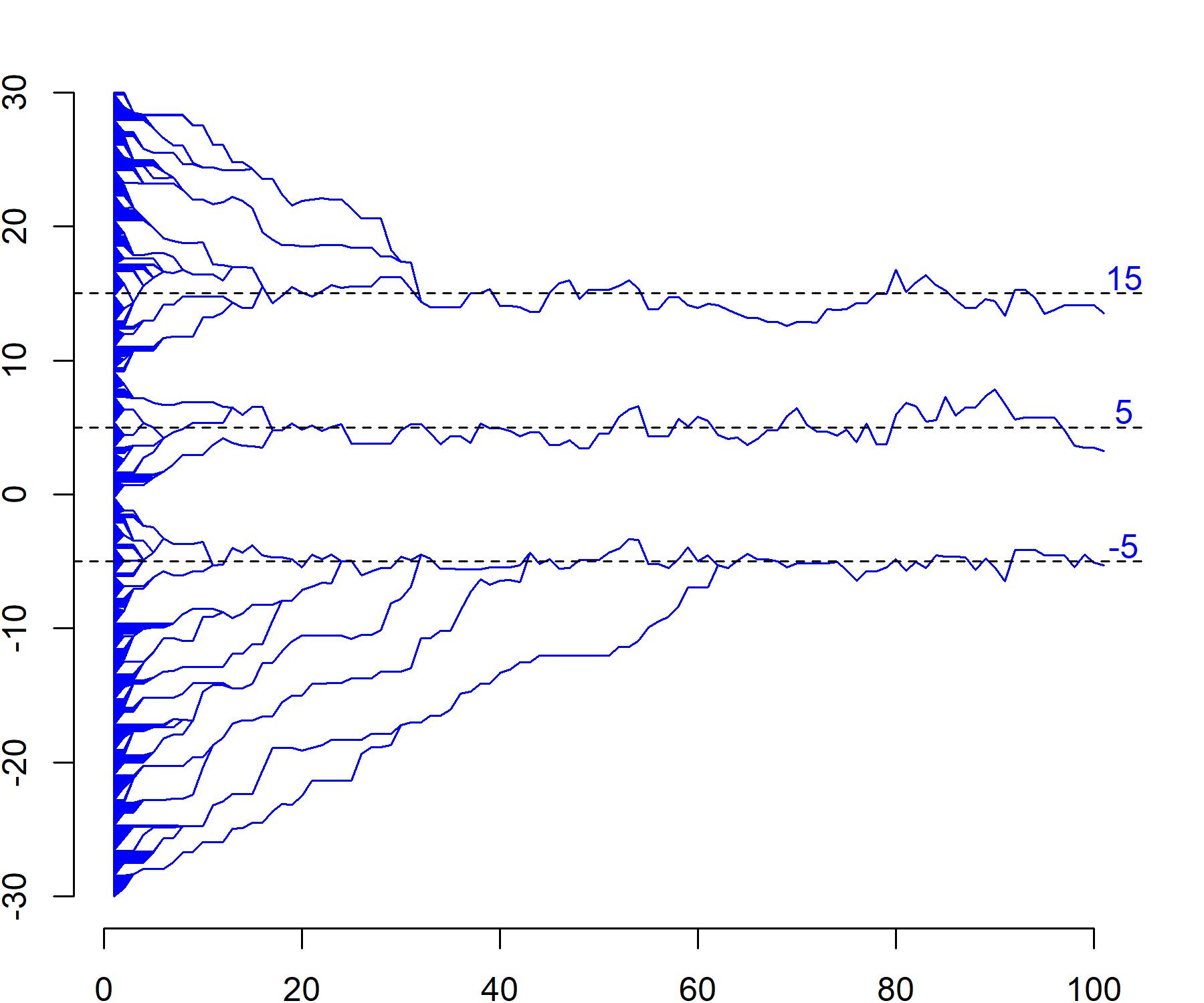}
\end{center}
\caption{Applying Metropolis-multishift simultaneously on $(-30,30)$ with target $0.2N(-5,1)+0.2N(5,1)+0.6N(15,1)$. \label{fig:11}}
\end{figure} 

\begin{figure}[ht]
\begin{center}
\includegraphics[width=3in]{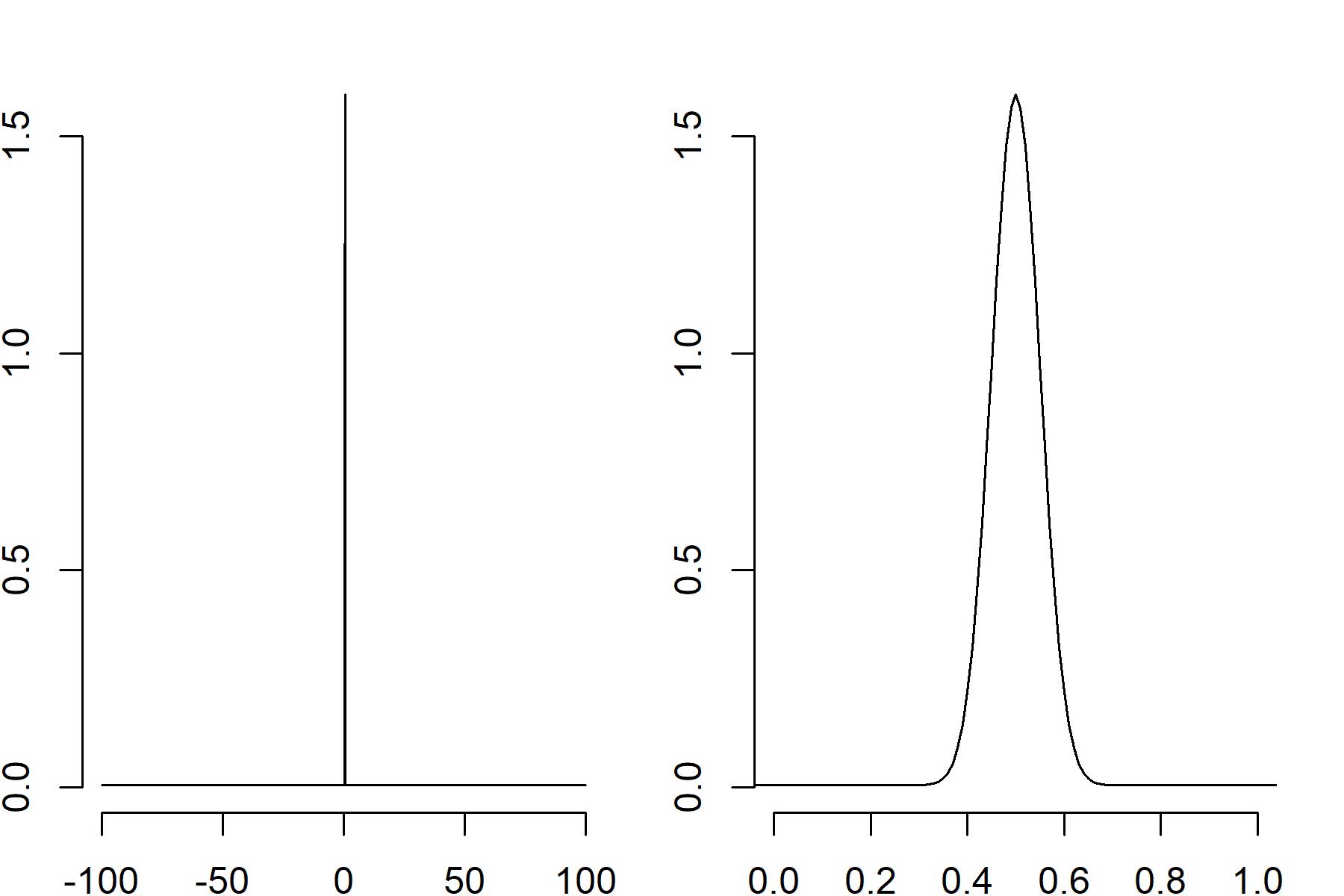}
\end{center}
\caption{Illustration of $0.8Uniform(-100,100)+0.2Beta(50,50)$ in far view of $(-100,100)$ looks like a single line and in close up $(0,1)$ it shows the density shape. \label{fig:13}}
\end{figure} 

\begin{figure}[ht]
\begin{center}
\includegraphics[width=3in]{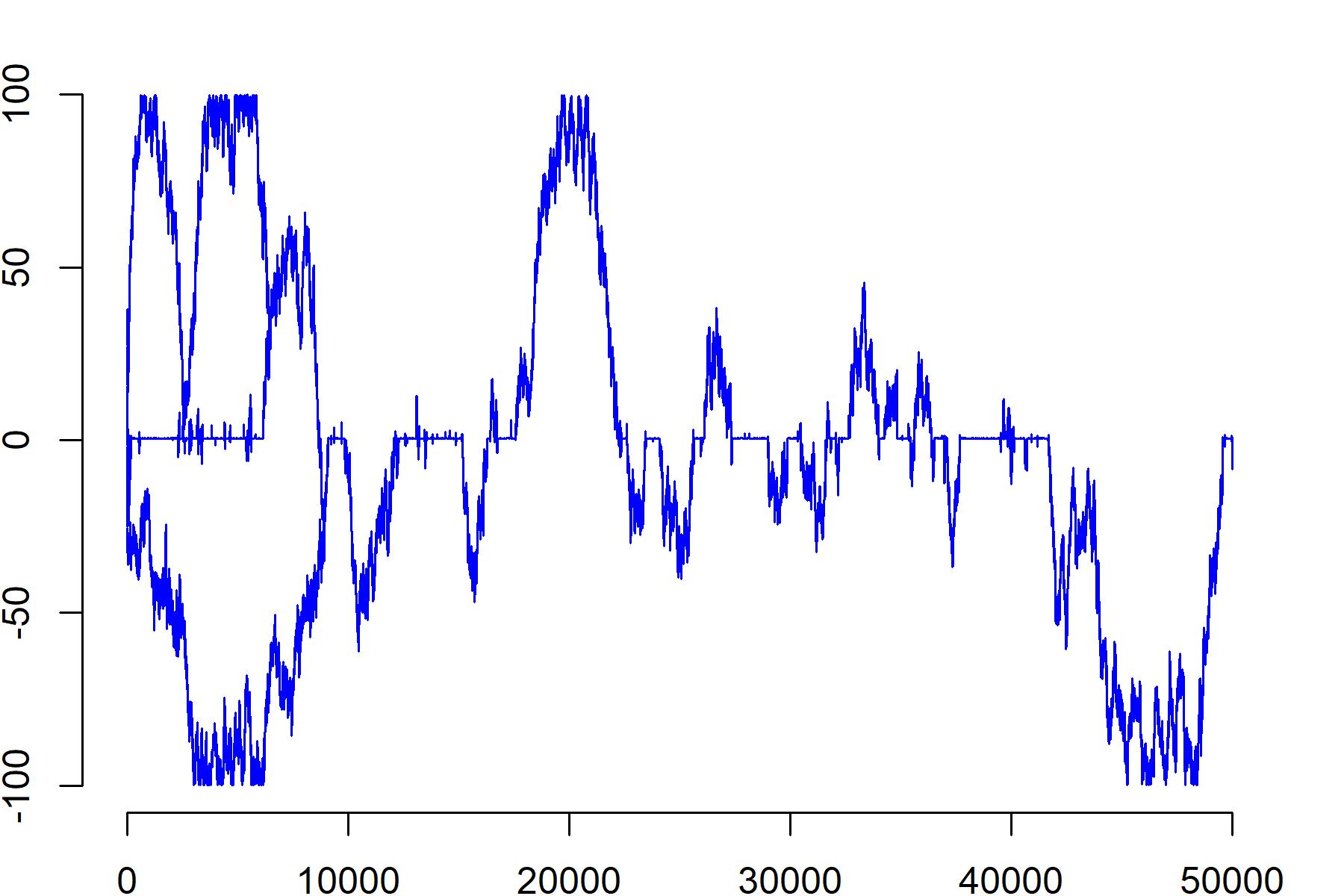}
\end{center}
\caption{Applying Metropolis-multishift simultaneously on range $(-100,100)$ with target $0.8Uniform(-100,100)+0.2Beta(50,50)$, and proposal multishift $N(0,1)$. \label{fig:14}}
\end{figure} 

\begin{figure}[ht]
\begin{center}
\includegraphics[width=3in]{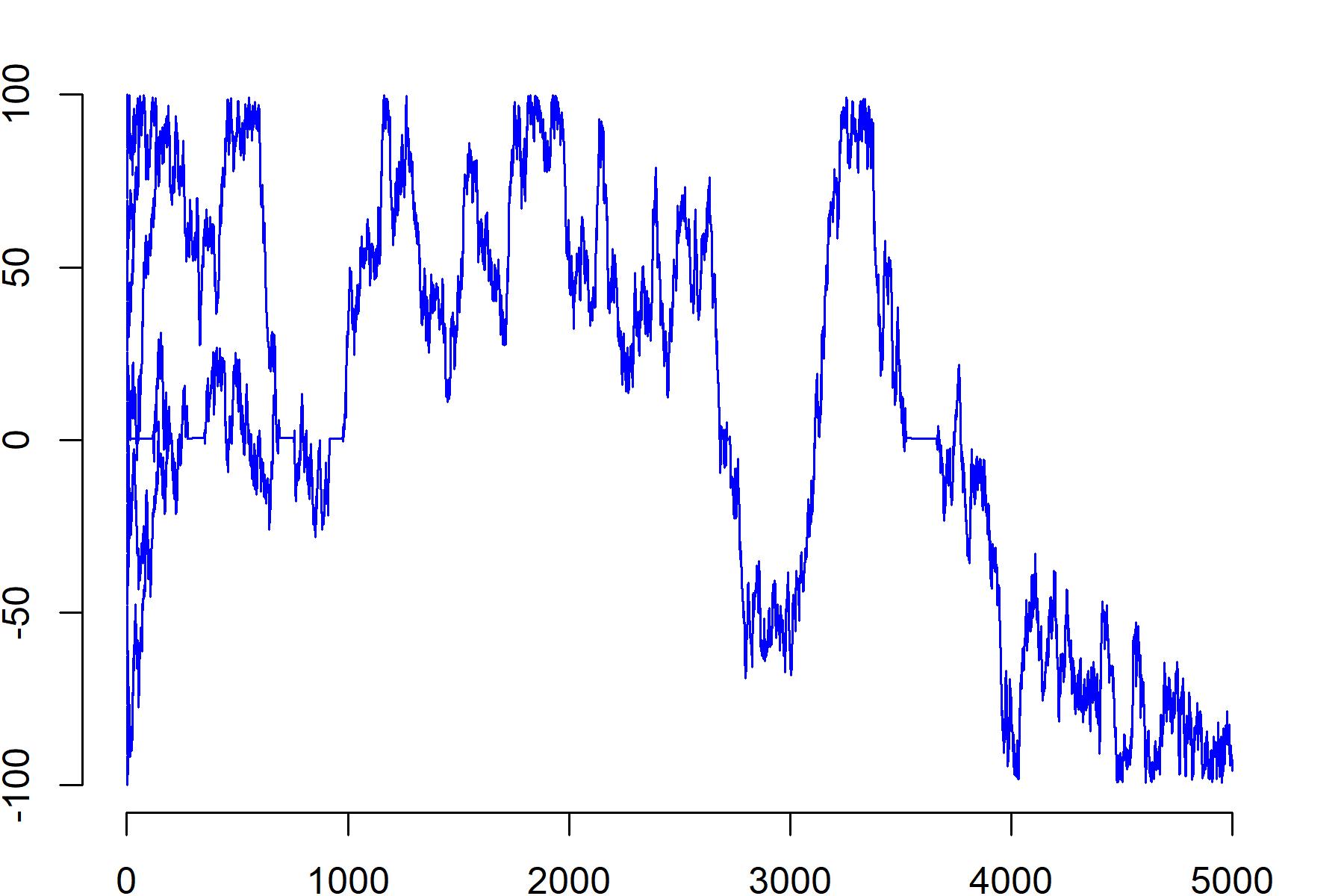}
\end{center}
\caption{Applying Metropolis-multishift simultaneously on range $(-100,100)$ with target $0.8Uniform(-100,100)+0.2Beta(50,50)$, and proposal multishift $N(0,3.5^2)$. \label{fig:15}}
\end{figure} 

\begin{figure}[ht]
\begin{center}
\includegraphics[width=3in]{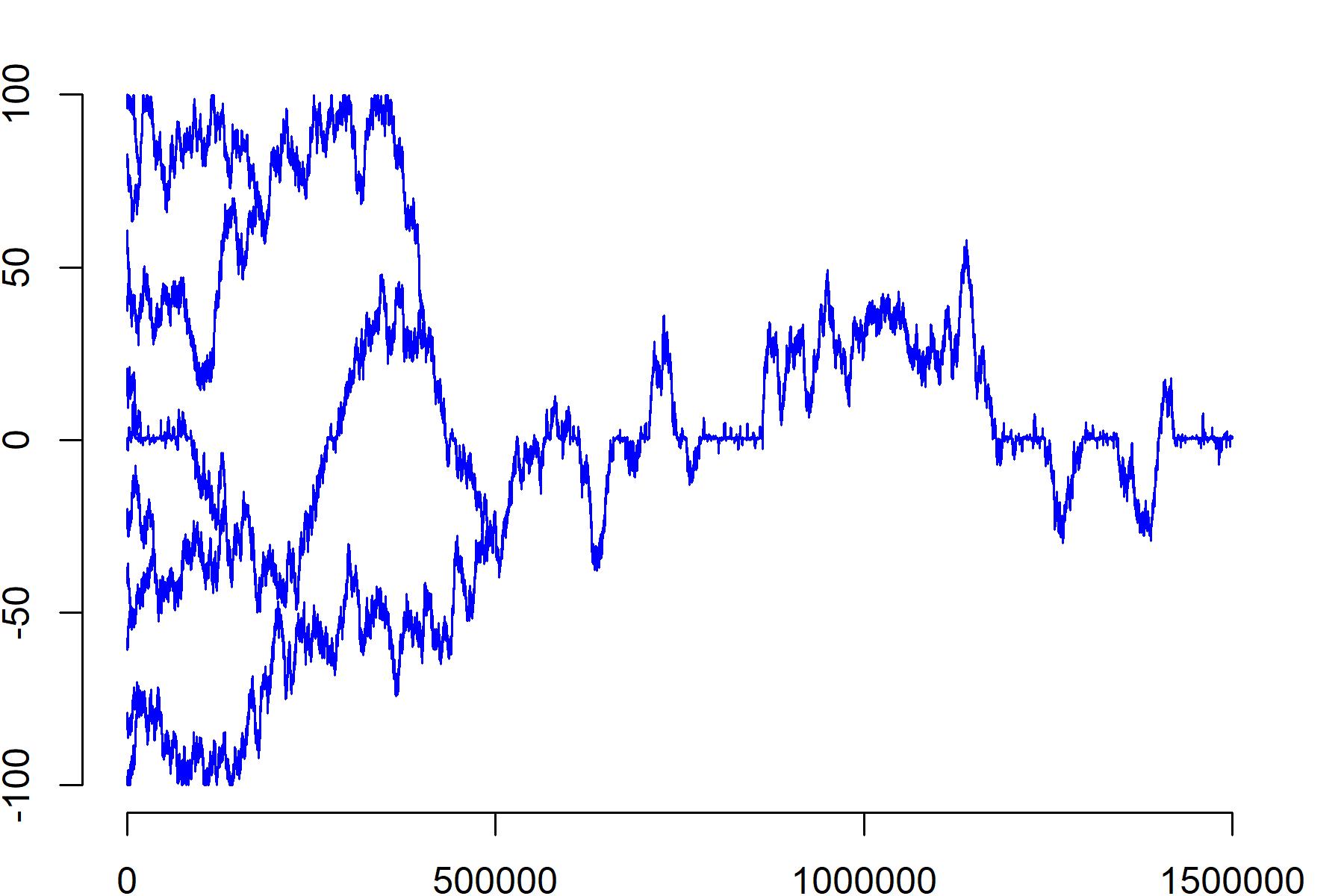}
\end{center}
\caption{Applying Metropolis-multishift simultaneously on eleven paths on the range $(-100,100)$ with target $0.9Uniform(-100,100)+0.1Beta(500,500)$, and proposal multishift $N(0,0.1^2)$. \label{fig:16}}
\end{figure} 

\begin{figure}[ht]
\begin{center}
\includegraphics[width=4.5in]{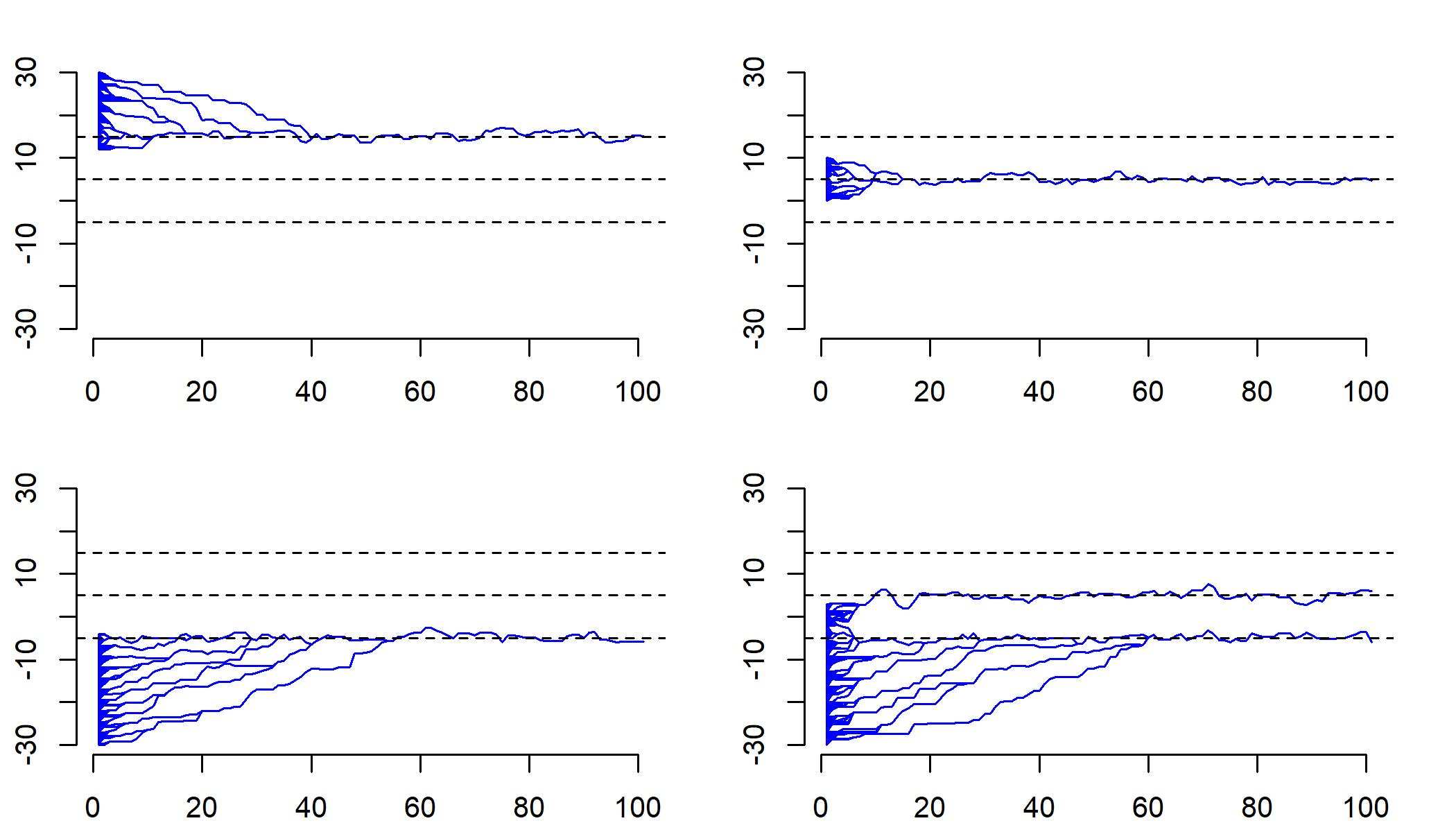}
\end{center}
\caption{Visual demonstration of the Metropolis-multishift algorithm applied simultaneously over different starting ranges with a target distribution of $0.2N(-5,1) + 0.2N(5,1) + 0.6N(15,1)$ using a proposal distribution of $N(0,1)$. \label{fig:17}}
\end{figure}

\begin{figure}[ht]
\begin{center}
\includegraphics[width=4.5in]{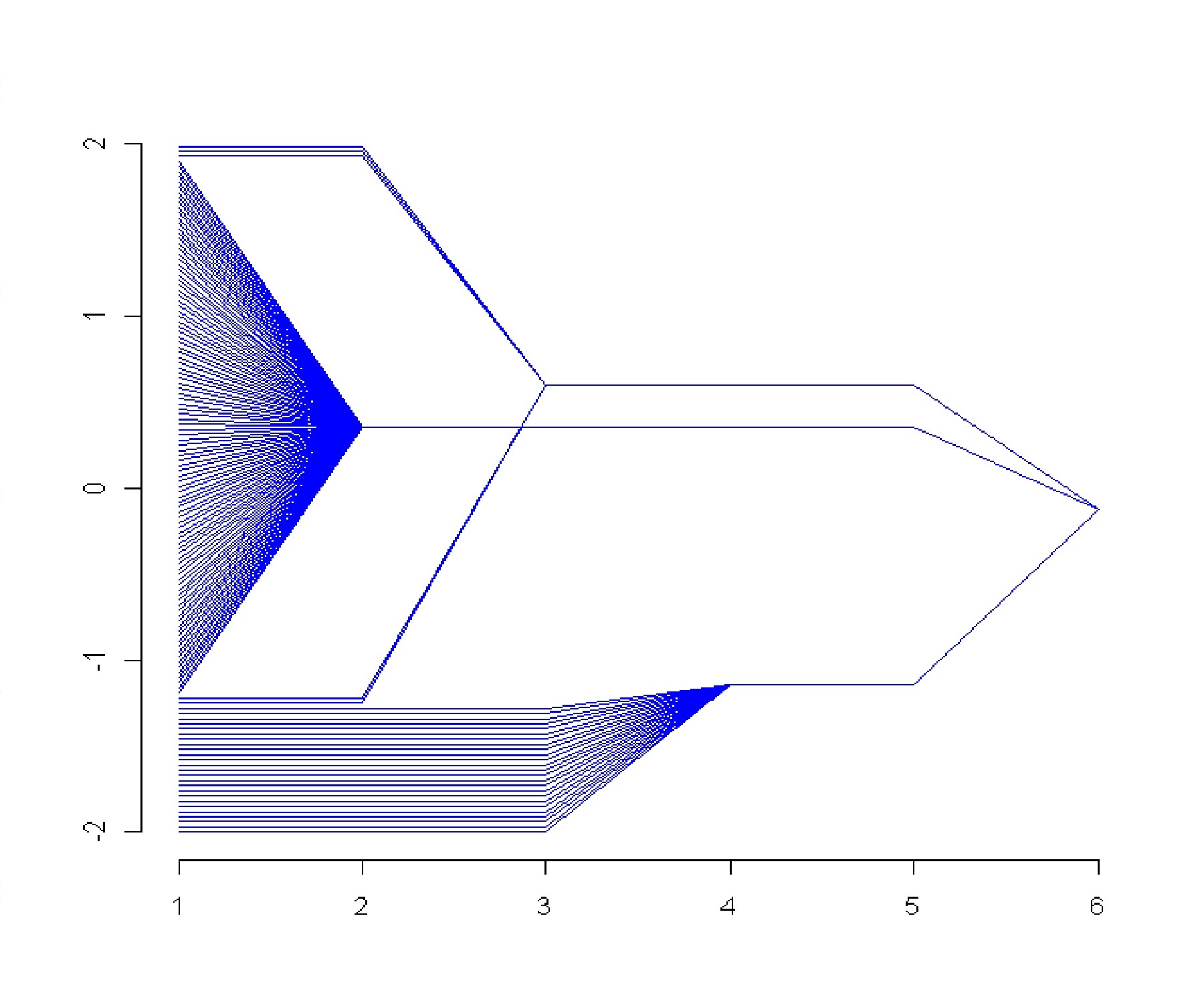}
\end{center}
\caption{Illustration of Metropolis-multishift coupler in the first six steps. \label{fig:19}}
\end{figure}

\begin{figure}[ht]
\begin{center}
\includegraphics[width=6.5in]{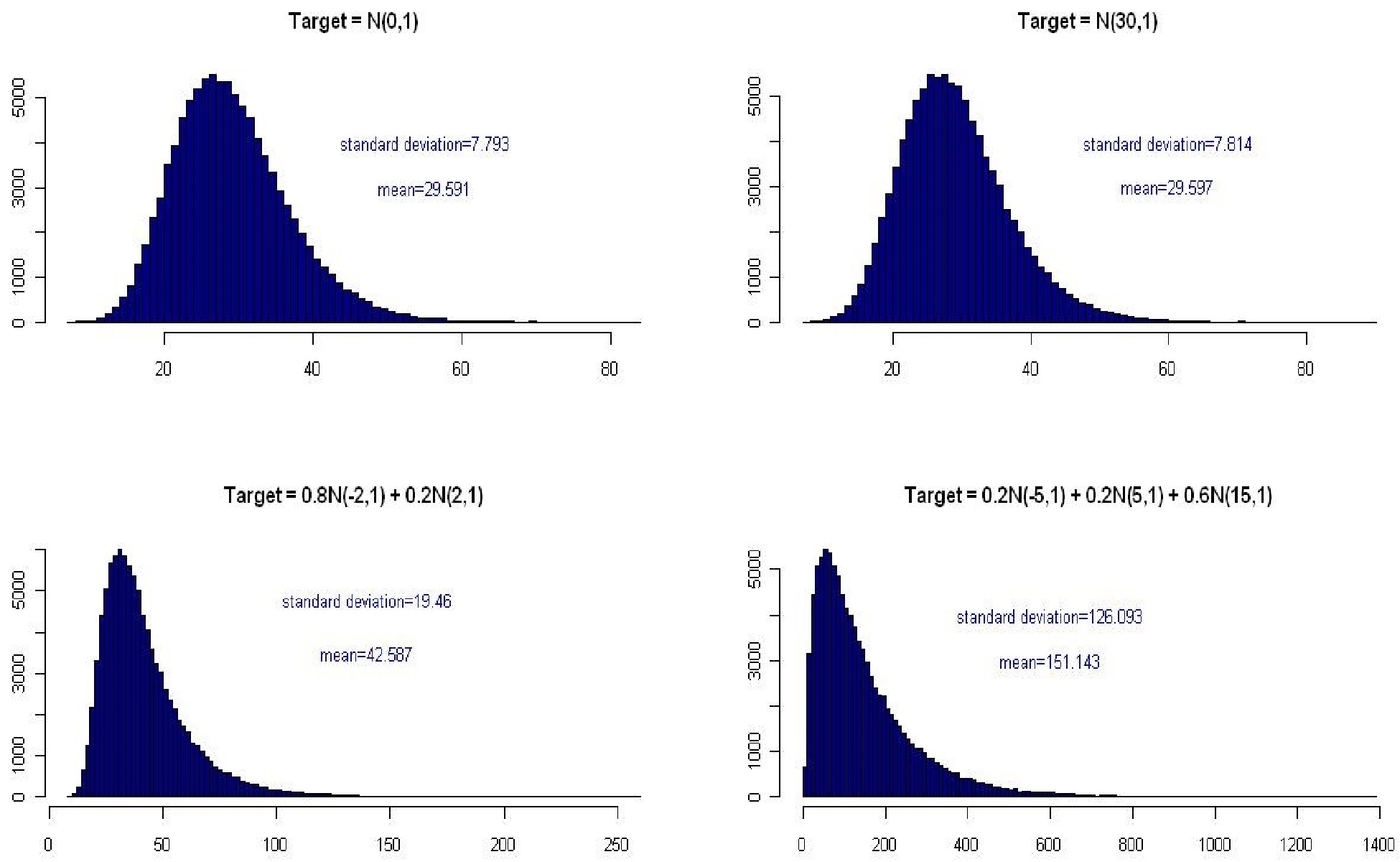}
\end{center}
\caption{Histogram of coalescence time with different targets. \label{fig:20}}
\end{figure}

\begin{figure}
\subfloat[N(0,1)]{\includegraphics[width = 3in]{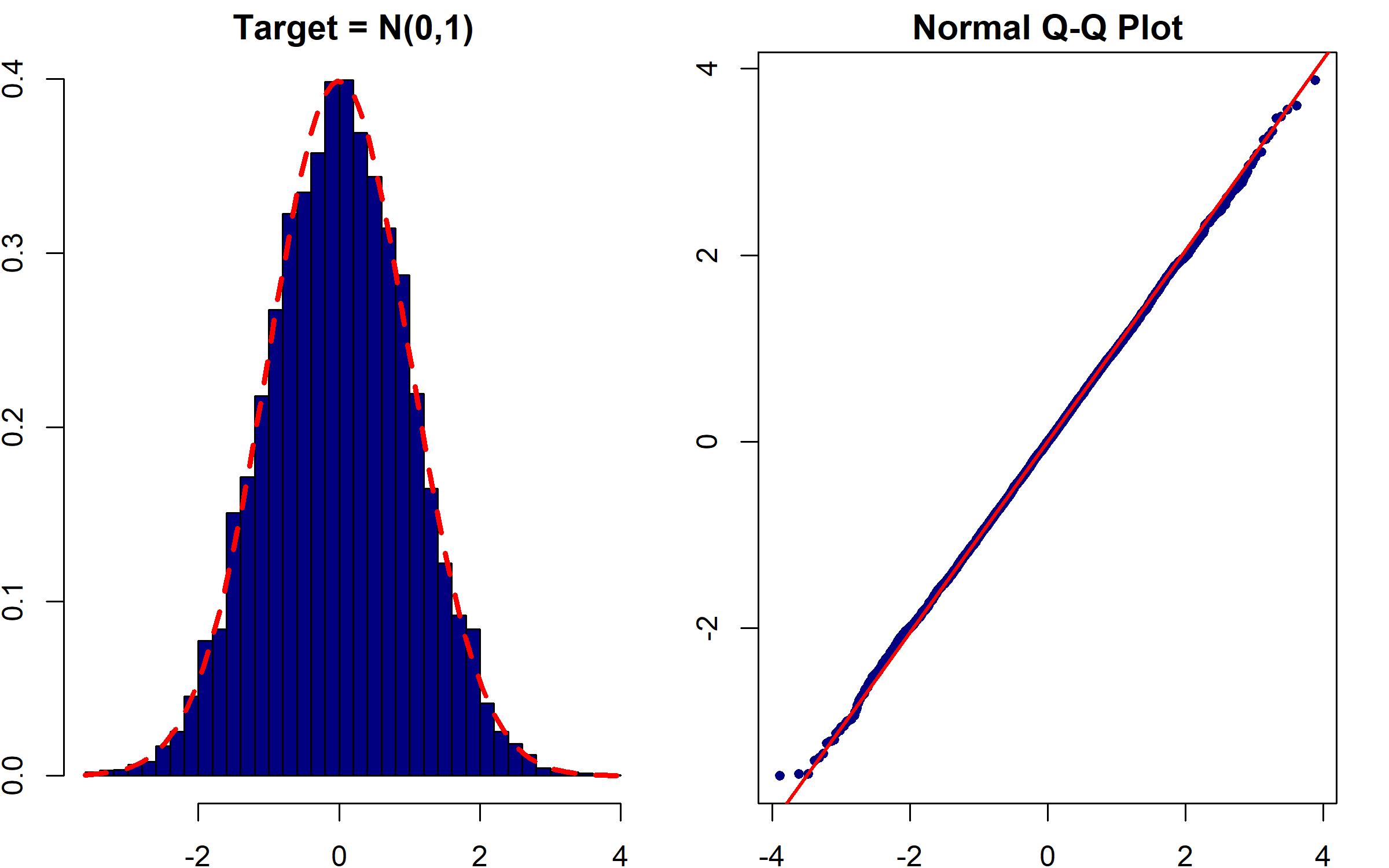}} 
\subfloat[N(30,1)]{\includegraphics[width = 3in]{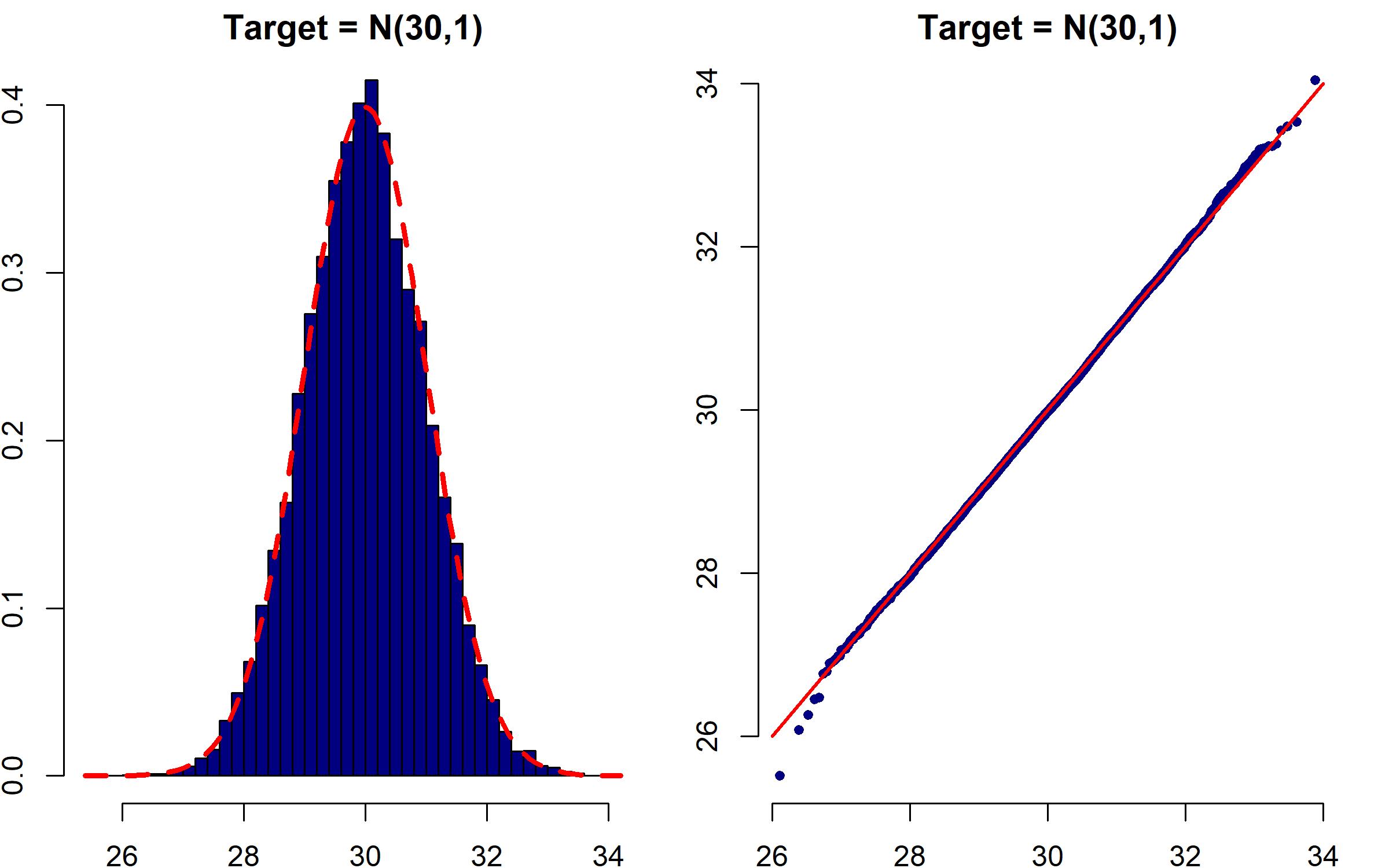}}\\
\subfloat[0.8N(-2,1)+0.2N(2,1)]{\includegraphics[width = 3in]{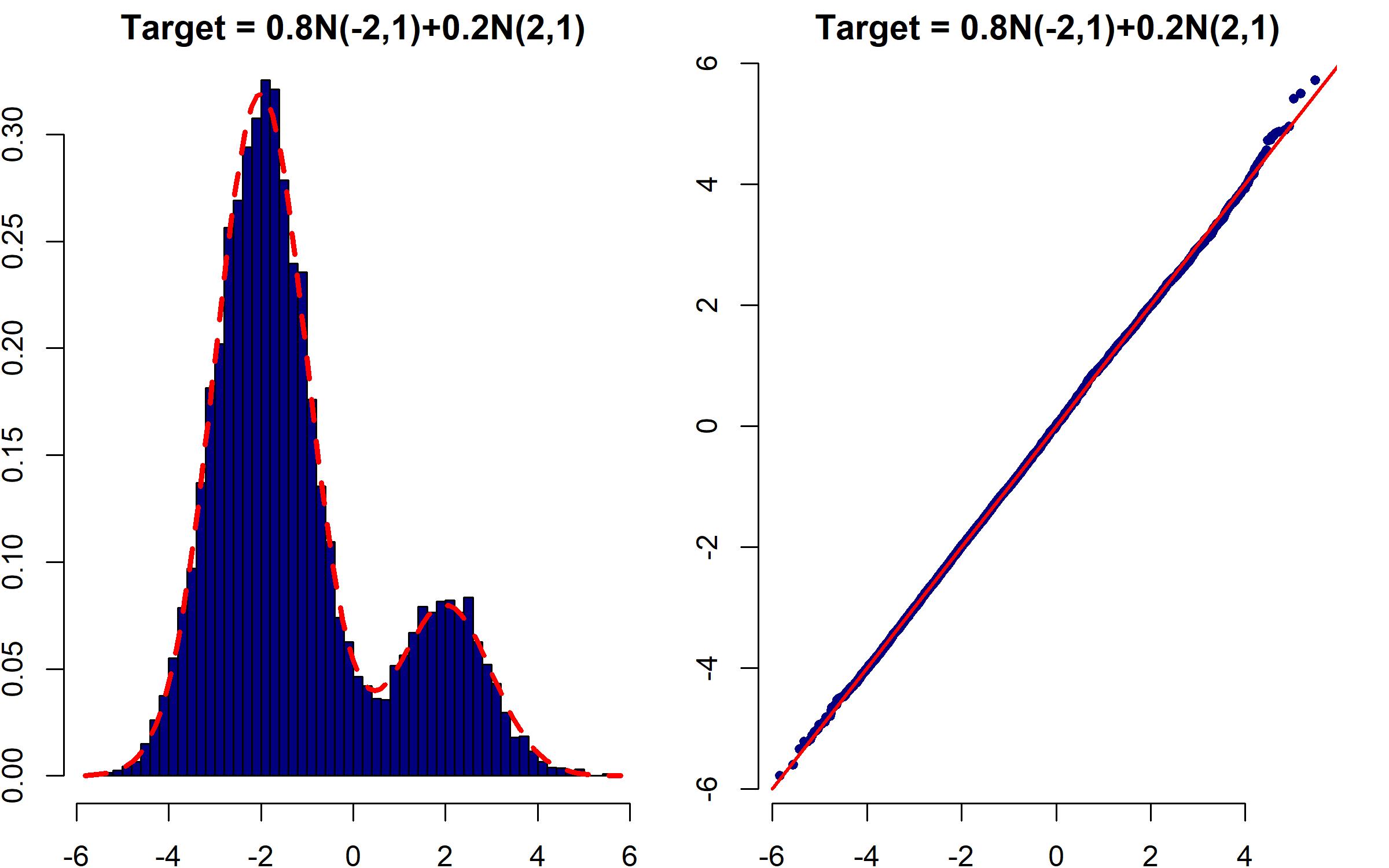}}
\subfloat[0.2N(-5,1)+0.2N(5,1)+0.6N(15,1)]{\includegraphics[width = 3in]{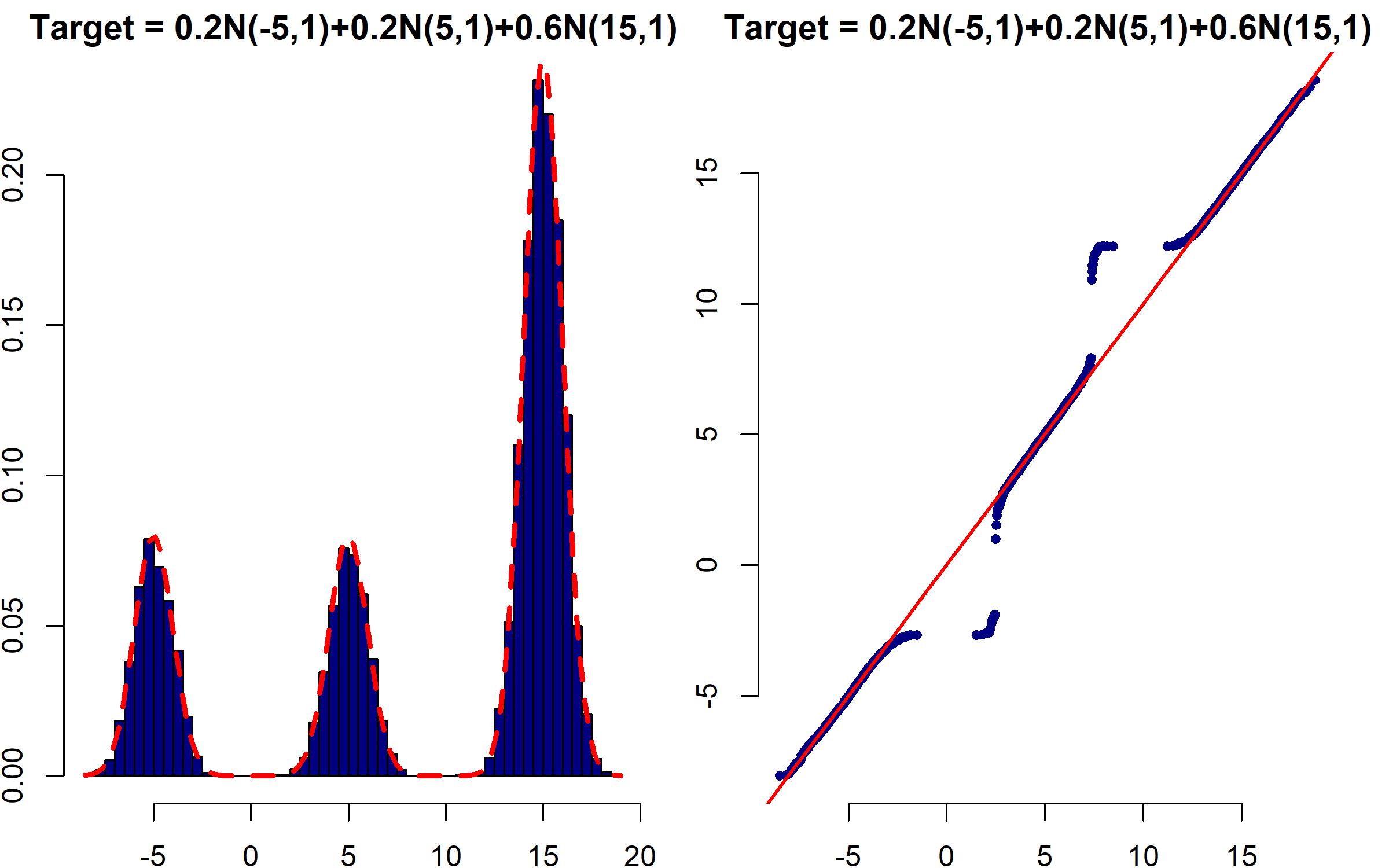}} 
\caption{Test of fitting generated samples with target distribution of Table \ref{T2}.}
\label{fig:22}
\end{figure}

\begin{figure}[ht]
\begin{center}
\includegraphics[width=6.5in]{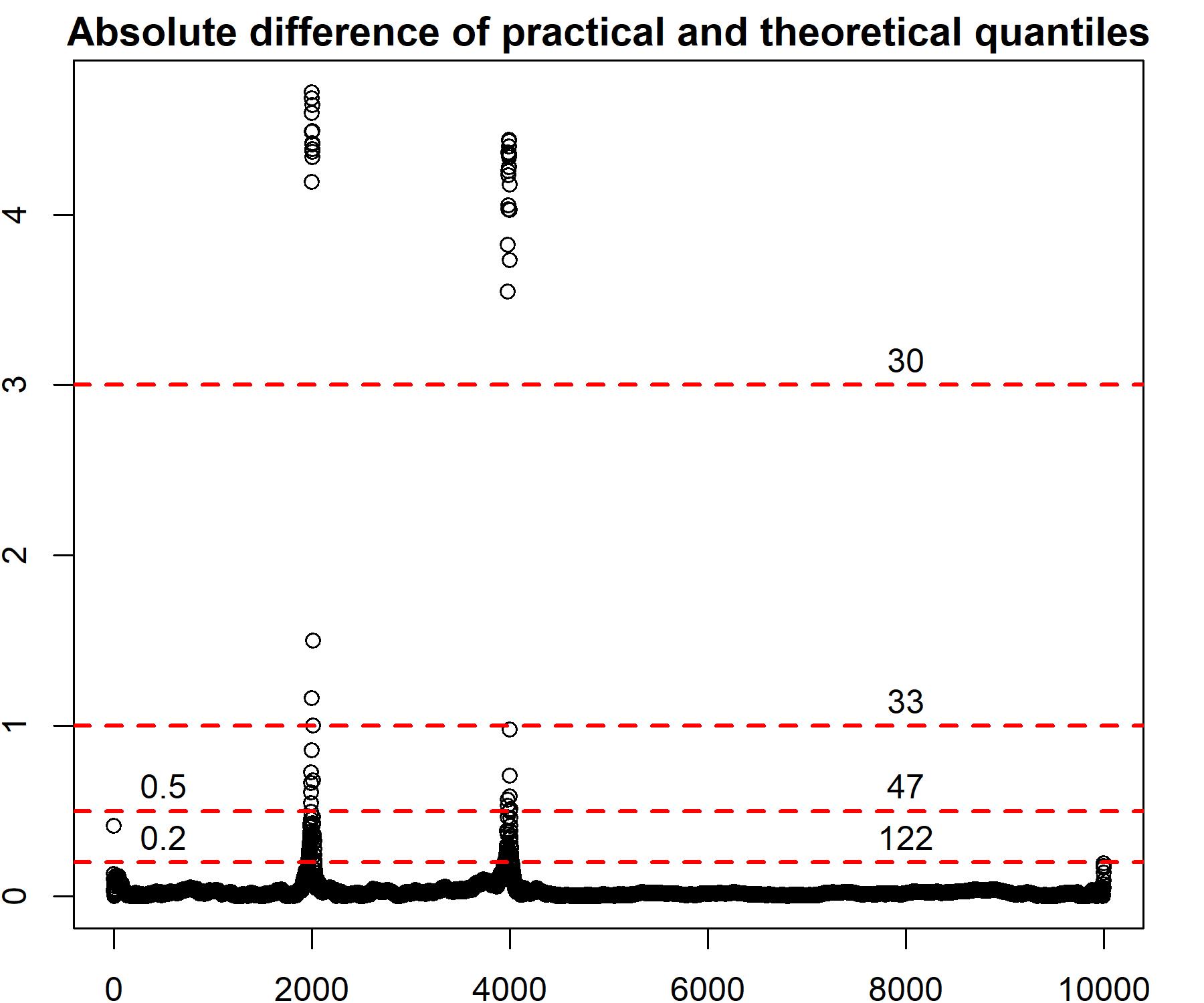}
\end{center}
\caption{The number of out fit points in Figure \ref{fig:22}. \label{fig:23}}
\end{figure}

\clearpage
\subsection{R package}
 R-package ROCFTP.MMS containing code to perform ROCFTP with Metropolis-multishift described in the article. The package also contains examples provided in the article. 
 \emph{https://cran.r-project.org/web/packages/ROCFTP.MMS/index.html}

\subsection{CFTP R code} \label{CFTP code}
\begin{verbatim}
ev<- function(vec)
{
even<- NULL
for(i in (1: floor(length(vec)/2)))
{
even<- c(even, 2*i)
}
return(even)
}

L.R<- function()
{
 z<- rnorm(1,0,1)
 u<- runif(1,0,dnorm(z))
main<- sqrt(-2*log(sqrt(2*pi)*u, base = exp(1)))
  L<- -main
  R<- main
  m<- c(L,R)
  return(m)
}

update<- function(ro,inits,m,x)
{
even<-ev(inits)
odd<- (1:length(inits))[-even]
f<- rep(0, length(inits))
for(i in even)
{
f[i]<-floor((inits[i]*ro+m[2]-x)/(m[2]-m[1]))*(m[2]-m[1])+x
}
for(j in odd)
{
f[j]<-floor((inits[j]*ro+m[2]-x)/(m[2]-m[1]))*(m[2]-m[1])+x
}
return(f)
}

multishift<- function(ro,inits,niter)
{
pred<-matrix(inits, nrow=1)
random<-NULL
for (i in 1:niter)
{
m<-L.R()
x<- runif(1,m[1],m[2])
g<- update(ro,pred[i,],m, x)
pred<-rbind(pred,g)
random<- rbind(random,c(m,x))
}
return(cbind(pred[2:nrow(pred),],random))
}

multishift1<- function(ro,inits,niter, random)
{
pred<-matrix(inits, nrow=1)
for (i in 1:niter)
{
m<-random[i,1:2]
x<- as.numeric(random[i,3])
g<- update(ro,pred[i,],m, x)
pred<-rbind(pred,g)
}
return(cbind(pred[2:nrow(pred),],random))
}

CFTP<- function (ro,start)
{
random<-NULL
inits<- start
niter<- 2
tempsamp<- multishift(ro,start,niter)
random<- tempsamp[,3:5]
tempsamp<- tempsamp[,1:2]
repeat{
niter<-2*niter
samples<- multishift(ro,start,niter)
random1<- samples[,3:5]
samples<- samples[,1:2]
cond<- samples[nrow(samples),]
inits<- samples[nrow(samples),]
b<- matrix(c(rep(NA,nrow(samples)*ncol(tempsamp)-2),start),
           ncol=ncol(tempsamp),byrow=TRUE)
sapmples<-cbind( b ,samples)
h<-multishift1(ro,inits,nrow(random), random)
h<- cbind(tempsamp,h[,1:2])
tempsamp<- rbind(sapmples,h)
random<- rbind(random1,random)
if(length(unique(cond))==1) break
}
b<- c(rep(NA, ncol(tempsamp)-2),start)
tempsamp<- rbind(b,tempsamp)
matplot(tempsamp, type="l", col="blue", lty=1, axes = FALSE, ylab=" ")
}

CFTP(0.92,c(-100,100))
\end{verbatim}

\end{document}